\documentclass[letterpaper,twocolumn,10pt]{article}
\usepackage{usenix,epsfig,endnotes}
\usepackage{booktabs}
\usepackage{multirow} 
\usepackage{xspace}
\usepackage{subfigure}
\usepackage{enumitem}
\usepackage{graphicx}
\usepackage{wrapfig}
\usepackage{csquotes}
\usepackage{xurl}
\usepackage{authblk}
\usepackage{amssymb}
\usepackage[normalem]{ulem}

\newcommand{\vrsocialapp}{multi-user VR applications\xspace}
\newcommand{\Vrsocialapp}{Multi-user VR applications\xspace}
\newcommand{\onevrsocialapp}{multi-user VR application\xspace}

\newcommand{\transform}{\textsc{Transform}\xspace}

\newcommand{\newpar}[1]{\vspace{0.05cm}\noindent\textbf{#1.}\xspace}
\newcommand{\newparagraph}[1]{\vspace{0.05cm}\noindent\textbf{#1}\xspace}
\newcommand\blfootnote[1]{%
  \begingroup
  \renewcommand\thefootnote{}\footnote{#1}%
  \addtocounter{footnote}{-1}%
  \endgroup
}

\begin{document}

\date{}

\title{\Large \bf Remote Keylogging Attacks in Multi-user VR Applications}

\author[1*]{Zihao Su}
\author[2*]{Kunlin Cai}
\author[1]{Reuben Beeler}
\author[1]{Lukas Dresel}
\author[1]{Allan Garcia}
\author[1]{Ilya Grishchenko}
\author[2]{Yuan Tian}
\author[1]{Christopher Kruegel}
\author[1]{Giovanni Vigna}
\affil[1]{University of California, Santa Barbara}
\affil[2]{University of California, Los Angeles}

\maketitle


\begin{abstract}
As Virtual Reality (VR) applications grow in popularity, they have bridged distances and brought users closer together. 
However, with this growth, there have been increasing concerns about security and privacy, especially related to the motion data used to create immersive experiences. 
In this study, we highlight a significant security threat in \vrsocialapp, which are applications that allow multiple users to interact with each other in the same virtual space. 
Specifically, we propose a remote attack that utilizes the avatar rendering information collected from an adversary's game clients to extract user-typed secrets like credit card information, passwords, or private conversations.
We do this by (1) extracting motion data from network packets, and (2) mapping motion data to keystroke entries. 
We conducted a user study to verify the attack's effectiveness, in which our attack successfully inferred 97.62\% of the keystrokes.
Besides, we performed an additional experiment to underline that our attack is practical, confirming its effectiveness even when (1) there are multiple users in a room, and (2) the attacker cannot see the victims. 
Moreover, we replicated our proposed attack on four applications to demonstrate the generalizability of the attack.
Lastly, we proposed a defense against the attack, which has been implemented by major players in the VR industry.
These results underscore the severity of the vulnerability and its potential impact on millions of VR social platform users.

\end{abstract}

\blfootnote{* Both authors contributed equally to this research.}

\section{Introduction}

As technology advances, Virtual Reality (VR) has gained significant attention and has become an easily accessible technology in people's lives. Experts estimated that over 171 million people use VR globally in 2024~\cite{vr24}. 
An emerging trend in VR is its use in multi-user applications~\cite{Insights}. 
These applications are becoming popular as they provide virtual spaces for users to interact, especially in situations in which environments dramatically improve the user experience.
VR applications are unique in their ability to translate users' movements in the real world into corresponding movements of their virtual avatars, thereby making users less cognizant of the gap between real and virtual worlds. 
\Vrsocialapp further extend the immersive experience of VR applications by accommodating various forms of communication and by rendering avatars for users across all application clients. 

Although using real-life motion data to render avatars across different clients can create a better immersive experience for users, it requires the transmission of motion data over the Internet. 
As mentioned by Nair et al.~\cite{nair2023truth}, user motion data is very sensitive and can be used to derive personal information such as the identity, anthropometric measurements (e.g., height and wingspan), as well as demographic details (e.g., age and gender) of users. 
Unfortunately, current \vrsocialapp do not offer adequate protection for motion data.
As a result, the beneficial functionality of this data in VR environments becomes a side channel that leaks users' private information. 


Recent research has demonstrated the feasibility of performing keylogging attacks (attacks that attempt to infer a user's keystrokes) against VR by leveraging side-channel motion information tied to typing behavior. 
For instance, VR-Spy~\cite{al2021vr} performs this attack by utilizing the side channel from channel state information of WiFi signals in the victim's local environment. 
HoloLogger~\cite{luo2022holologger} and TyPose~\cite{slocumgoing} employ malware to harvest hand or head tracking data from the victim's device to predict the victim's keystrokes. 
Another approach by Zhang et al.~\cite{zhang2023s} capitalizes on the side channel sourced from rendering performance counters in VR devices to infer user-typed numbers. 
A key assumption made by all these attacks is the ability to install malware or a dedicated surveillance implant in a user's {\it local} environment in order to obtain local motion-related data. 

In our work, we propose a keylogging attack that also leverages motion data, but it operates under a significantly less stringent assumption. 
That is, adversaries are able to execute our keylogging attack remotely.
The only requirement is that they need to be in the same virtual room as the victim. 
This assumption makes our attack more practical and makes all users of \vrsocialapp potential victims.

Adapting keylogging attacks to remote contexts while achieving high performance can be challenging. 
Unlike local keylogging attacks that leverage local sensors, we choose to recover typing-related motion directly from network packets received by the adversary's client.
Although this motion data is sent to all users in the virtual room via network packets, the applications themselves operate as black boxes. 
This creates difficulties in understanding how this information is transmitted to each remote application client and in which format. 
Furthermore, reversing the packet encoding or semantics can be difficult due to the lack of tools to debug the applications at runtime, given that these apps are frequently protected by anti-cheating engines and DRM (digital rights management) components. 
These difficulties collectively make recovering motion data from network packets a challenging task.
Moreover, even if we manage to successfully recover motion data from the packets, it is uncertain whether the motion data in the packets is high-quality-enough for an accurate keylogging attack. 
This is because applications often use methods such as compression and under-sampling to ensure stable and efficient network transmission. 
Consequently, the remote motion data has lower fidelity compared to the original sensor data. 
In this paper, we have overcome these challenges with our proposed attack, which can accurately reconstruct the motion data required to infer the keystrokes.

Our attack consists of four steps, each extracting more fine-grained information about the typing activity from the previous step.
In these four steps, we aim to understand and accurately leverage the data in network packets, converting the data to recover keystrokes and thereby executing the remote keylogging attack. 
We evaluated our attack through a user study conducted on Rec Room~\cite{recroom}, one of the most widely used \vrsocialapp with more than 15 million users~\cite{recroomuser}.
We successfully reconstructed the typed secrets at top-1 accuracy of \textbf{97.62\%}. 
Our attack results demonstrate that we can infer almost all user-typed information correctly, even though remote settings offer lower-fidelity information compared to local settings. 
Furthermore, we performed an additional experiment in which (1) there are multiple users in the room, and (2) the attacker does not see any other user (from their application client's point of view).
Our attack achieves comparable performance under this setting (top-1 accuracy of \textbf{97.53\%}), demonstrating the practicality of our attack.
Lastly, we replicated our attack on three additional applications and performed user studies on them, in which we achieved comparable performance across all applications (top-1 accuracy of 98.24\%, 98.27\%, and 99.07\% respectively from the additional applications).
This result further demonstrates that our attack is generalizable across applications.
We reported our attack to Rec Room, the three additional applications, SteamVR, and Unity.
The developers of Rec Room, Sing Together: VR Karaoke~\cite{sing_together} and SteamVR~\cite{steamvr} acknowledged the issue and Rec Room also awarded us a bounty for the vulnerability.
Furthermore, SteamVR and Rec Room have created patches to defend against our proposed attack, following our defense suggestions as outlined in Section~\ref{sec:discussion}. 

\newparagraph{Contributions:}

(1) To the best of our knowledge, we are the first to demonstrate the feasibility of remote keylogging attacks in the context of \vrsocialapp. 
Our approach enables more practical attacks under a remote setting, significantly enhancing the practicability and stealthiness compared to existing methods.

(2) We introduce a novel attack approach to overcome the challenge of recovering typing-related motion from a remote application client.
Additionally, we provided new tools for VR motion data processing, such as precise motion input control and precise cursor/keyboard measurement. 

(3) 
We also introduce an alternative attack strategy effective on network packets that 
have been partially reverse-engineered, by applying machine learning techniques to the raw bytes we extracted from the packets.
Our result demonstrates that a remote keylogging attack with only minimal manual reverse engineering effort is also possible, with trade-offs in additional attack setup and accuracy.


(4) We conducted user studies to assess the efficacy, practicality, and generalizability of our attack. 
In Rec Room, we analyzed typing data from 22,092 clicks (involving letters, numbers, and special characters) provided by 20 participants, as well as typing data from additional 2,431 clicks by two participants typing concurrently, while the attacker cannot see them (in the VR space).
We further analyzed typing data from 7,656 clicks provided by 9 participants for three additional applications (three participants for each application).
From our analysis, we provide insights into which real-world keystrokes are vulnerable to our attack. 
Also, from the feedback of the participants, we studied users' awareness and concerns about our attack to understand its implications.

(5) We propose countermeasures for \vrsocialapp to mitigate our remote keylogging attacks and avoid privacy leakage associated with transmitting motion data across the Internet. Rec Room and SteamVR have adopted and implemented these countermeasures.


\section{Background}

\newpar{Multi-user VR Applications}
As the name suggests, a \onevrsocialapp provides interaction opportunities for users, allowing them to communicate with other players across the Internet. 
Typically, users can communicate by both typing and speaking. 
The typing functions are usually achieved using a virtual keyboard, either in-application or through overlay applications such as Steam Chat or Messenger. 
These newly emerging services allow users to interact in realistic digital worlds, and they distinguish themselves from traditional social networking services by capturing and translating real-world user motions into their virtual landscapes~\cite{wei2022communication}. 

\begin{figure}[h]
\centering
\includegraphics[width=0.2\textwidth]{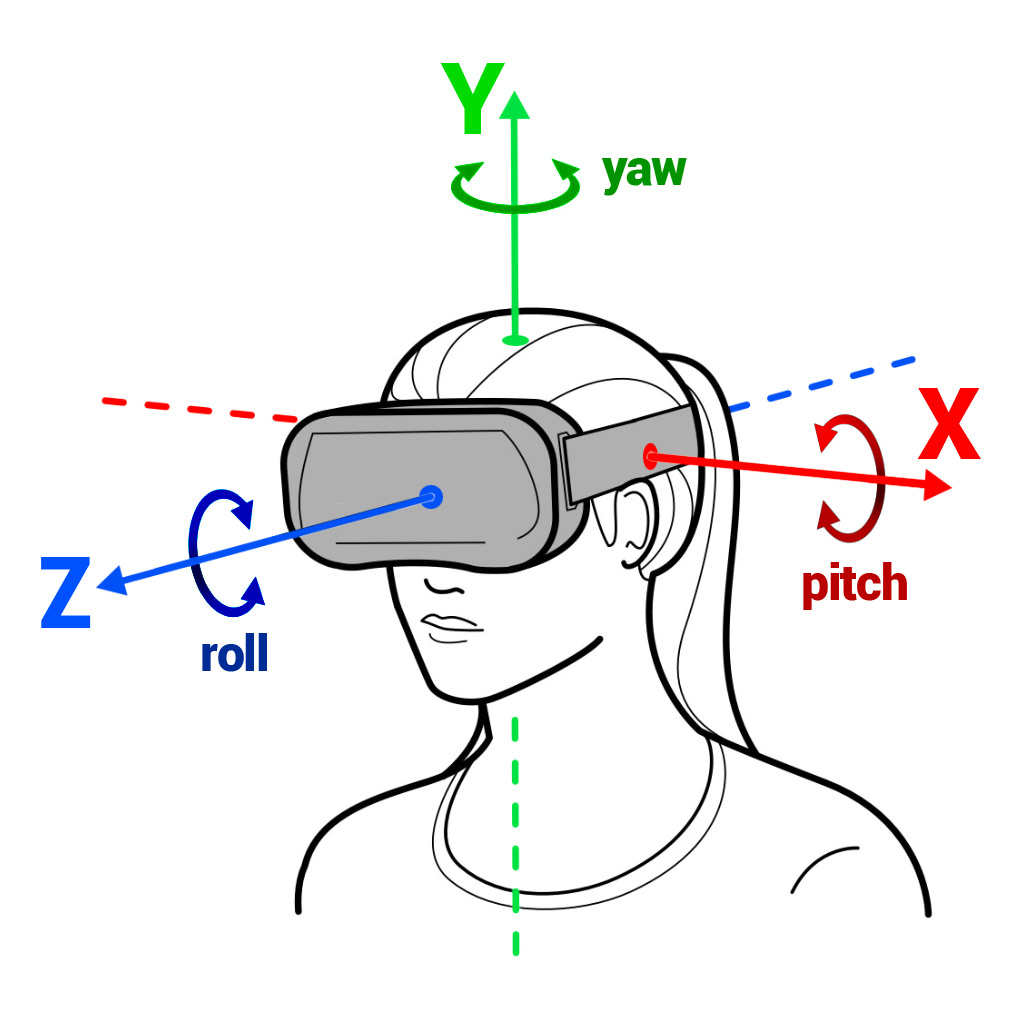}
\caption{Example of user head motion in 6DOF~\cite{sixdof}.}
\label{fig:6dof}
\end{figure}

\newpar{Motion Update in Multi-user VR Applications} 
Motion, which has six degrees of freedom (6DOF) as shown in Figure~\ref{fig:6dof}, is normally described by position (x, y, z) and rotation (roll, pitch, yaw). 
In this paper, we define \transform to be the composition of position and rotation.

Popular game engines describe motion with their own data structures that encapsulate \transform.
For example, the Unity Engine, the most popular game engine for \vrsocialapp~\cite{unity_market}, has a standard data structure to represent \transform using  a Quaternion (for rotation) and a Vector3 (for position)~\cite{unity_transform}.
In Unity-based applications, this standard data structure is transmitted over the network to synchronize user motions.

\label{motion_update}

\begin{figure}[h]
\centering
\includegraphics[width=0.48\textwidth]{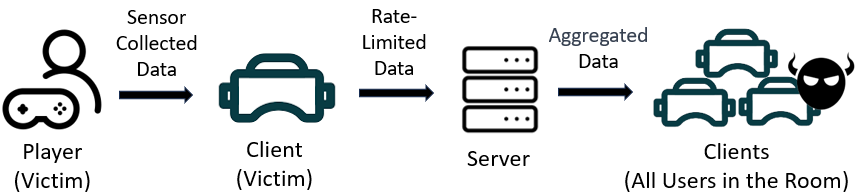}
\caption{Motion data flow in \vrsocialapp.}
\label{fig:motionflow}
\end{figure}

For \vrsocialapp, 
the server continuously receives motion data updates from clients and broadcasts them to all clients~\cite{Photon_Engine}, as seen in Figure~\ref{fig:motionflow}.
Typically, the synchronization process employs UDP packets with established libraries such as Photon PUN/Fusion~\cite{Photon_Engine}, Unity Netcode \cite{netcode}, and Mirror~\cite{mirror} to ensure stable and accurate motion transmission. 
However, the motion data may not be fully identical after the transmission, as sometimes, applications (e.g., Rec Room) may use lossy compression on the motion data update, trading off a slight loss of fidelity of motion data in exchange for a lower burden on the transmission process.


Moreover, the transmitted motion data is also under-sampled from the original motion data. 
In current multi-user applications, the server usually sends updates to the clients at a rate of 20-30 packets per second. This rate takes into account the network bandwidth limitation on both the server and client sides, ensuring efficient communication without overloading either end~\cite{networkrate}.
But applications usually run at 60-120 frames per second (FPS)~\cite{Linde_2022}, so they have adapted mechanisms like interpolation~\cite{interpolation} to compensate for performance loss.

\newpar{VR Typing Mechanism}
\label{click_mechanism}
In VR, there are multiple options for text input. 
The most commonly deployed methods include using voice or typing on a virtual keyboard~\cite{dudley2019performance}. 
In this paper, our primary focus is on the typing method that utilizes a virtual keyboard. 
Typing on a virtual keyboard involves two major steps: (1) moving the cursor, which can be a controller cursor or a virtual hand, to the target key on the virtual keyboard, and (2) selecting the key by performing a click, either by pressing a button on the controller or by poking at the key with a hand gesture. 
Since both of these steps involve the movement of the user's hand, \vrsocialapp map the corresponding motion to the avatar. This ensures the avatar feels realistic and accurately represents the details of the user's hand motion.
\begin{figure}[ht]
\centering
\includegraphics[width=0.43\textwidth]{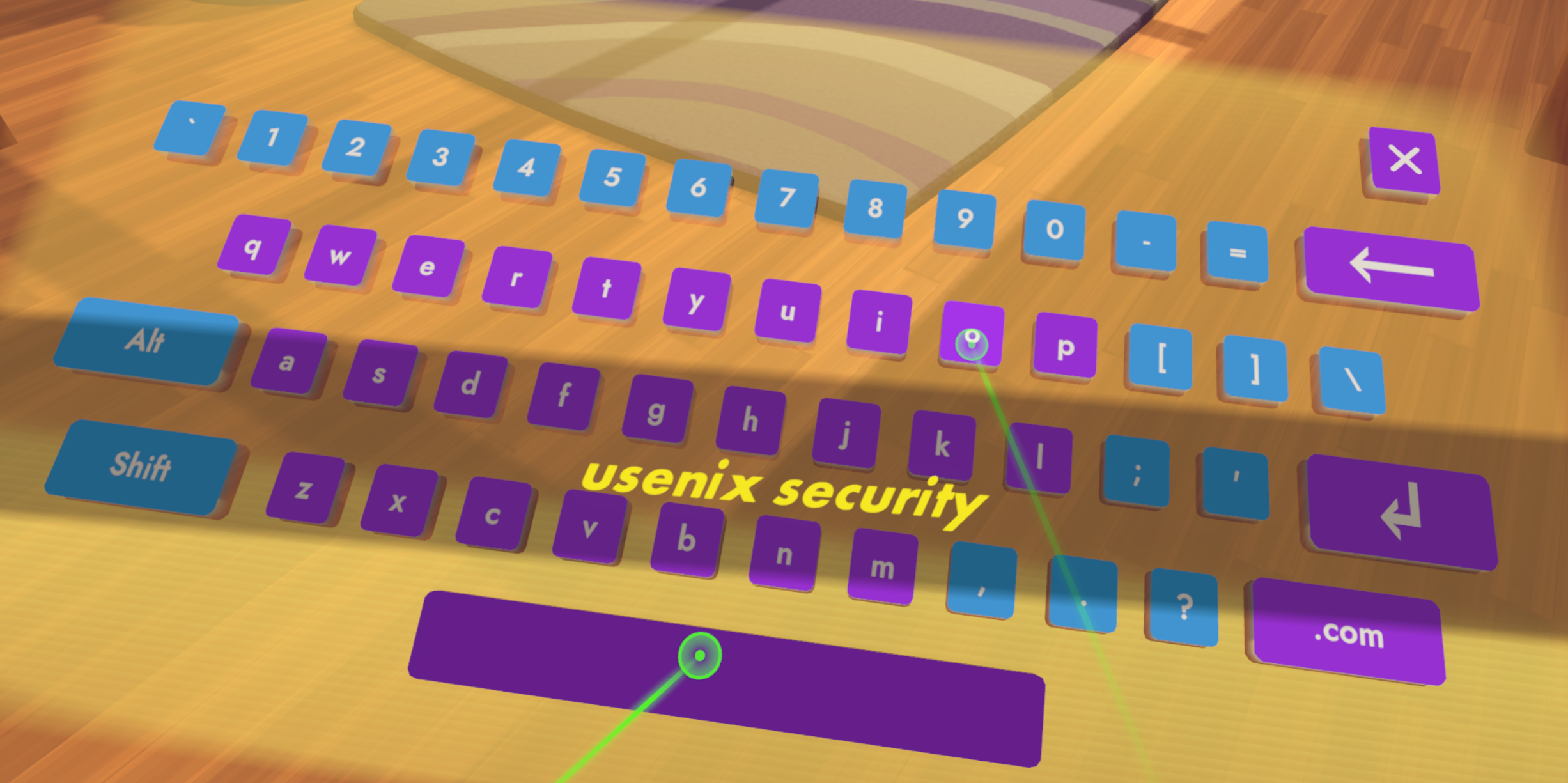}
\caption{Example of a participant typing in Rec Room using a virtual keyboard.}
\label{fig:typing}
\end{figure}

For the click mechanism, we focus on clicks represented by button presses on a controller, which is a widely adopted method for both interaction between users and typing on virtual keyboards. 
Considering the importance of click functionality, Unity offers standardized APIs, such as OVR~\cite{ovr} and Input system~\cite{xrinput}, to help applications detect these clicks. 
These APIs use the trigger value (i.e., how deep the trigger button is pressed down, from 0.0 to 1.0) to detect clicks.
Typically, the API registers a click when the trigger value is higher than a specific threshold. 
Since the trigger button is typically held for more than one frame per click, to avoid repeated firing, the first frame where the trigger is pressed is recognized as the click moment~\cite{inputsystem}.

\section{Threat Model}
\label{threatmodel}
\newpar{Adversary Objective}
Our attack aims to extract user-typed secrets inside \vrsocialapp.
By extracting these secrets, an attacker may gain access to the following types of sensitive information:


(1) \textit{Credit Card Information:}
Modern VR applications often incorporate payment gateways to facilitate transactions, requiring users to input sensitive financial details, such as credit card numbers.
Such information represents a profitable target for adversaries aiming for unauthorized financial access.
Conventionally, credit card numbers are sequences of digits.

(2) \textit{User Authentication Data:}
Within the context of \vrsocialapp, authentication mechanisms, such as password inputs, are employed when users attempt to access their accounts, private virtual rooms, or private virtual assets. 
Adversaries can actively seek these credentials to gain unauthorized access to user accounts.
These credentials typically consist of alphanumeric characters, and may sometimes include special characters.

(3) \textit{Private Conversation:} \Vrsocialapp incorporate social elements that include both professional and social activities. 
Users often engage in private chats within these applications to communicate with business partners or friends, which could include the exchange of sensitive information, such as business-related data or personal matters. 
The private conversation is usually composed of strings of alphabetical characters.

Given that recent applications offer functionalities such as in-app purchases of virtual items and private user chats, the entry of private information, as explained above, has become common in multi-user VR applications.

\newpar{Adversary Knowledge}
For the scope of the attacks discussed in this paper, we operate under the assumption that the adversary operates a remote client (an unmodified client located on the adversary's side that can receive updates about other players sent by the server) 
of a \onevrsocialapp and does not have access to the victim's devices or local environment. 
The target \onevrsocialapp should have both typing capabilities and the functionality to synchronize user motion across a network infrastructure. 
To the best of our knowledge, these features are ubiquitously supported across current \vrsocialapp.

Also, we assume that an adversary is a legitimate user (by downloading the application and registering an account in the application) who can enter a virtual room with other users. 
This presence makes the other users in the room potential victims (i.e., the attacker does not necessarily follow and target a specific user).
This assumption is reasonable because, in multi-user VR applications like Rec Room, most rooms are public to facilitate the applications' purposes of socializing and meeting strangers.
Being in the same virtual room allows the adversary's application client to receive motion updates from all other users in the room, including their typing-related motion data, even if their avatars are not visually seen by the adversary in the application.
Moreover, the attacker can differentiate and group motion updates from different victim users.
This is because current network protocols for multi-user VR applications require a unique user identifier 
to associate user avatars with their network updates.
Therefore, by grouping motion updates using this user identifier, the attacker can analyze motion updates from different user entities, and perform the attack independently on each of them, thereby stealing keystrokes from all users in the room.
Note that this user identifier only allows the attacker to group the inferred keystrokes and associate these keystrokes to any network updates linked to the user identifier (e.g., in-game username); further linking such information back to each user's identity in the real world is out of scope.


Additionally, given that the adversaries control their own clients, we assume they have the ability to access the binary files of the application client and capture the network traffic that their client sends and receives.
Furthermore, we assume that the adversary can prepare for the attack by studying the application behavior.
For example, the attacker can create multiple accounts, perform typing-related experiments on their own clients using these accounts, and observe the visual outputs of the application (e.g., how the keyboard looks).

\section{Approach}
\label{sec:approach}

The basic insight behind our approach is that \vrsocialapp transmit a user's VR motion data over the network and make it available to all other users who are in the same virtual room.
This includes the motions that occur when a user is typing. 
Thus, typing-related motions are received by everyone in the same room.
Consequently, anyone can analyze the network traffic and reconstruct the user's typing motion to infer the user's keystrokes.

\begin{figure}[ht]
\centering
\includegraphics[width=0.47\textwidth]{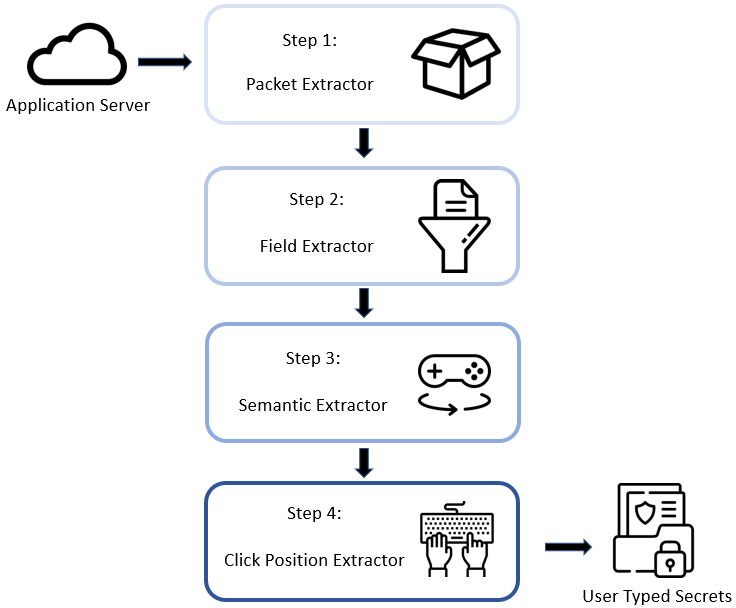}
\caption{Our approach consists of four extraction steps to extract keystrokes from network traffic.}
\label{sys_overview}

\end{figure}

As mentioned previously, it is challenging to extract motion data from packets and perform accurate keylogging attacks using low-fidelity motion data. 
We solve these challenges with a four-step approach, which is outlined in Figure~\ref{sys_overview}. 
The first step takes the network packets and filters for those including motion data. 
It is followed by a step that parses the packets and identifies the data fields within each packet. 
The third step recovers the semantics of these fields and extracts the motion data. 
In the last step, the extracted motion data is mapped to key positions, which provides us with the prediction of the user's keystrokes. 
Our approach differs from existing keylogging approaches as it only requires network packets collected from an attacker's own application client as inputs. 


In this section, we present how we implemented each of these steps for a specific target. 
In particular, we choose Rec Room, one of the most popular \vrsocialapp that has more than 15 million users~\cite{recroomuser},
as an example to demonstrate our attack.




\subsection{Step 1: Packet Extraction}
\label{subsec:packet}

The first step of the approach is to capture the raw network traffic and filter out the packets irrelevant to motion data.
To recover motion data, 
we use Wireshark~\cite{wireshark} and locate the incoming network traffic of our Rec Room client by its application port.
Note that a single VR application is typically communicating with different servers (IP addresses) that handle different channels (such as voice data, messages, motion updates, etc.), but each server uses a fixed IP address while we stay in the room.
In order to find the server that sends traffic specifically related to motion data, we group the network traffic from the VR application by source IP address, and take the packets from the IP address with the highest transmission frequency.
These packets will most likely contain the motion data sent from Rec Room's server, as it updates motion data with clients at a high frequency (a fixed rate of 15 packets per second). 
Therefore, by performing this filtering step, we can exclude the majority of network traffic irrelevant to motion data, and limit the scope of analysis for the next step. 

For other \vrsocialapp, this strategy can be similarly applied, as the application needs to update users' motion data at a high frequency to ensure smooth motion (see Section~\ref{motion_update} for details about motion updates for \vrsocialapp), which makes it easy to isolate the motion update traffic. 
In addition, we can also use other common characteristics of motion updates to further eliminate irrelevant packets (e.g., motion updates are usually sent as UDP packets). 

\subsection{Step 2: Field Extraction}
\label{subsec:field_extraction}
Once we obtain the packets that contain the user's motion data, we need to parse the packets and extract data fields serialized within the packets, so that we can extract the semantics from the fields later. 
In Rec Room, we implemented this step by first using a parser for Photon Engine~\cite{Photon_Engine}, a networking library that Rec Room uses and whose network protocol parser can be found in an open-sourced project~\cite{pundissector}. 
With this parser, we are able to parse the packet into data fields, and extract the types (e.g., integer, float, etc.) and values of the data fields (see Appendix~\ref{appendix:example_photon_packet} for an example packet parsed with a generic Photon protocol parser). 
However, Photon allows developers to define and transmit their own custom objects (i.e., objects of custom data type, which may include multiple fields defined by the developers), and Rec Room uses this feature.
Without knowing the fields serialized inside of these objects, we only see them as blobs of raw bytes.
To parse these objects, we decompile Rec Room and find how it defines and deserializes the relevant objects (see Appendix~\ref{reverse_engineer_appendix} for our insights in performing this step), which allows us to build a parser that breaks down these objects into individual fields. 

For other \vrsocialapp, we can replicate this implementation with mostly automated steps, and some reversing effort. 
It is likely that we can reuse the network protocol parser. Many applications use well-established networking libraries (e.g., Photon Engine and Unity Netcode~\cite{netcode}) instead of implementing networking functionalities from scratch.
For example, among the \textbf{34} \vrsocialapp that we investigated, \textbf{21} of them use Photon Engine, including Rec Room and VRChat \cite{vrchat}, two of the most popular \vrsocialapp on the market. 
Therefore, we can spend a one-time effort to find or implement the protocol parser and reuse it on other \vrsocialapp, and only spend reversing effort to recover the custom objects if the applications define them (as Rec Room does).
However, if it is a rare case that the application does not use any networking library, we will need to reverse how the application deserializes the entire packet, which requires more manual effort.

One possible extra obstacle in this step is that the network traffic may be encrypted. 
While it is rare that VR applications encrypt the packets that include motion data (only \textbf{1} out of the \textbf{34} \vrsocialapp that we investigated encrypts their packets that include motion data), this obstacle can still be overcome.
The key exchange process and decryption happen on the attacker's own client.
Therefore, the attacker can decrypt the network traffic with a user CA certificate to conduct a Man-in-the-Middle attack on their own client, and previous work~\cite{trimananda2022ovrseen} has successfully performed this step and decrypted network traffic for 140 VR applications. 



\subsection{Step 3: Semantics Extraction}

After parsing all the fields, it remains unknown which fields correspond to the typing-related information, and hence, we need to associate the fields with their semantics. 
Specifically, we aim to find the fields with the following semantics, which are needed for the next step:

(1) \textit{Body motion data}, which includes $\transform$ (i.e., the position and rotation) of the left hand, right hand, and head.
They are used to track a user's typing motion. 
Such data is widely used in  {\vrsocialapp} to update a user's body position.

(2) \textit{Click data}, which includes the trigger values of the left controller and right controller. They are used to determine when a user is performing a click. 
This data is typically included in  {\vrsocialapp} to update users' hand gestures.

(3)  \textit{Keyboard-opening event data}, which is signaled through the menu-opening event, is used to determine the keyboard position and when the user may start typing. 
This event data is specific to certain applications, but it is optional, as previous work \cite{gopal2023hidden} demonstrates that the keyboard location can also be accurately approximated using the bounds of hand motion data.
Also, the starting point of user typing can be inferred using click patterns as discussed in~\cite{wu2023privacy}.

(4)  \textit{User identifier}, which associates the motion update with the user (avatar) who that update belongs to.
This field allows us to extract a separate motion data stream for each user when there are multiple users in the room.

To identify fields (1) to (3), we perform an experiment in which we (a) programmatically provide controlled motion data by running simulated VR hardware inputs, (b) compute the changes in packet fields, and (c) manually observe how changes to the inputs correspond to changes in the packet fields.
This experiment is explained in detail in Section~\ref{precise_input_control}. Identifying field (4) is trivial since each motion update is annotated with a user identifier (a built-in field of the Photon protocol).

This step can also be replicated in other \vrsocialapp.
Since most \vrsocialapp use Unity, which uses a standard data structure to represent motion data (see Section~\ref{motion_update} for more details), we observed similar associations between the motion data inputs and the packet fields across applications.
Furthermore, this experiment can be performed on any other \onevrsocialapp, since it only relies on manipulating the VR hardware inputs and observing packet data.
Moreover, even if the mapping is not obvious (e.g., if an application applies obfuscation in the packets), we can resort to reversing the semantics from the application's binary by tracing how the packet fields are being used, which would require some more manual effort.

\subsection{Step 4: Click Position Extraction}
\label{subsec:clickpos}
Once we extract the motion data, the last step is to infer keystrokes by 
(1) finding the \transform (i.e., the position and rotation) of all keys, (2)
detecting the timing of the clicks, (3) calculating the cursor's \transform (the cursor refers to a point that projects a line to select keys) at click time, and (4) finding the intersection of the cursor's projection and keys. 
Essentially, these steps are reversing and emulating how multi-user VR applications compute keystrokes.

Firstly, we need to calculate the keys' \transform. 
To do this, we first need to know (a) how the keyboard is positioned relative to a user (see \textbf{Keyboard Measurement} in Section~\ref{precise_measurement} for how this is measured), and (b) how the keys are positioned on the keyboard (see \textbf{Key Measurement} in Section~\ref{precise_measurement}), both of which only need to be measured once for an application in the attack preparation stage, as they are fixed. 
Then, we can find the packet that contains the \textit{keyboard-opening event}, and use the \textit{body motion data} in the packet to find the user's \transform at the time of keyboard opening.
With this information, we can use the pre-measured values in (a) to calculate the absolute position of the keyboard, and use the pre-measured values in (b) to calculate the positions of each key. 

Secondly, we need to determine the timing of the clicks. 
That is, we find in which packet each click happens, matching one packet to a click, so that we can later analyze the motion data in each of these packets to understand where the cursor is at the time of the click. 
This can be done by analyzing the \textit{click data} and performing click detection (see Section~\ref{click_mechanism} for details about how clicks are performed and detected in VR). 

Thirdly, we need to calculate the cursor's \transform. 
To do this, we first need to know how the cursor is positioned relative to the hand, which is a fixed relation (e.g., the hand position is at the palm of the hand, whereas the cursor is at the tip of the index finger).
Similar to the pre-measured values of the keyboard, this also needs to be measured only once for an application (see \textbf{Cursor Measurement} in Section~\ref{precise_measurement}).
Then, for each packet matched to a click, we find the hand's \transform from the \textit{body motion data} and use this pre-measured value to obtain the cursor's \transform. 

\begin{figure}[h]
\centering
\includegraphics[width=0.38\textwidth]{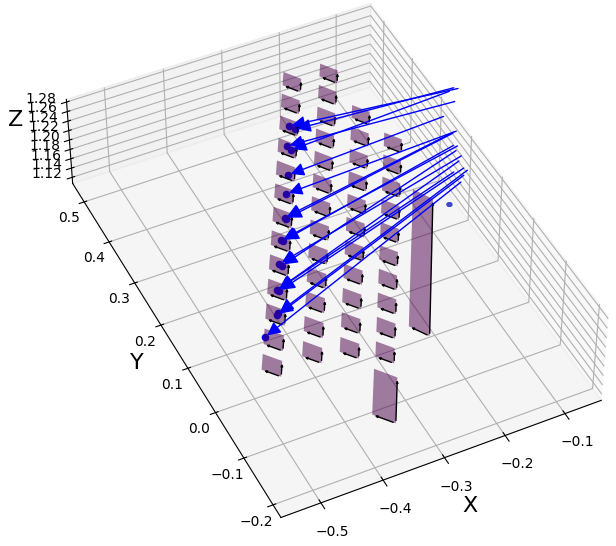}
\caption{By visualizing the extension of the cursor with blue arrow lines, we calculate its intersection with the keyboard, as indicated by the blue dots on the keys, to log the key selections.}
\label{fig:keyboard_visualization}
\end{figure}

In the fourth phase, knowing the cursor's and the keys' \transform during a click, we can detect the actual key that was clicked. 
Specifically, as illustrated in Figure~\ref{fig:keyboard_visualization}, we project a line (the blue arrow lines in the figure) following the cursor, and when this line intersects with any of the keys on the keyboard, we log this information (the blue dots on the keys in the figure). 
Intuitively, one can visualize a line through the fingertip of the avatar’s hand and see where this line points to and intersects the virtual keyboard. 
Note that this process is the same as how applications would calculate user keystrokes on local clients, and we are essentially ``replaying'' the keystroke entries extracted from the network packages in this step.

In other \vrsocialapp, the pre-measured values of the cursor and keyboard may be different from Rec Room, and we need to repeat this one-time measurement process. 
Note that this process can be performed on any other \vrsocialapp, as it operates by manipulating the VR hardware data and observing visual outputs.

\subsection{Reverse Engineering Challenges}
\label{perfect_motion}

In our approach, there is information that needs to be extracted once per application in the attack preparation stage: the semantics of key fields in network packets (Step~3) and the pre-measured values of the cursor and keyboard (Step~4). 
One of the most effective ways to obtain such information is to debug an application by accessing the application's memory while manipulating the inputs and observing the outputs at runtime. 
However, due to protections from anti-cheat engines~\cite{easy_anticheat,faceit,battleye}, attaching a debugger to \vrsocialapp is difficult. 
In this section, we introduce two methods to overcome these challenges without the need to bypass the anti-cheat engines. 
They employ only the readily available channels: VR hardware input, network packet output, and visual output (as mentioned in our threat model in Section~\ref{threatmodel}).
These methods enable us to (1) understand the associations between hardware inputs and packet fields by providing controlled inputs
 and (2) precisely measure the cursor and keyboard \transform (i.e., position and rotation) using geometric manipulations of the hardware inputs and visual validation.


\subsubsection{Precise Input Control}
\label{precise_input_control}

In Step~3, we aim to recover the semantics of packet fields. 
To do that, we want to provide precisely controlled inputs to the application and observe how changes are reflected in the fields of observed packets.
VR device inputs are highly dimensional (e.g., a controller has 6DOF inputs to represent its position and rotation), and we need to isolate and control each input dimension to understand the effect of their changes.
Moreover, such inputs cannot be easily controlled manually because of the sensitive sensors built into the VR hardware. 
For example, if we want to determine the effect of only changing the $x$ coordinate of the left controller position, we need to (1) fix the right controller and head-mounted display (i.e., the VR headset) and avoid even the slightest movements, (2) move the left controller along the $x$-axis without any movement in the $y$-axis and $z$-axis, and (3) prevent any rotation in any directions. 
Using the actual controller,  performing such a task ``by hand'' is essentially impossible.

To solve this challenge, we leverage Nvidia's VR Capture and Replay (VCR) tool~\cite{vcr}, which allows developers to record tracking data from VR devices and replay them. 
Specifically, we utilize the ``replay'' component of VCR to serve as a VR hardware simulator and use it to run programmatic inputs. 
With this feature, we can (1) locate the fields affected by each input by isolating changes in the input, and (2) uncover conversions from inputs to fields. 

First, we associate inputs and fields by observing their correspondence. 
For example, if we want to locate the $x$ coordinate of the left hand, we can (1) fix all other input components in our replay script and only change the $x$ coordinate, then (2) only fix the $x$ coordinate and change every other input component. 
If there is a field in the network packets that changes if and only if $x$ changes, it is highly likely that this field corresponds to $x$. 

Second, after we locate an input component's corresponding field, we can reason how they are converted by observing how the field value changes with the input. 
For example, we can vary the $x$ coordinate from -1.0 to 1.0 and see how the field for $x$ changes accordingly.

\subsubsection{Precise Measurement}
\label{precise_measurement}
In Step~4, we can parse the \transform (i.e., position and rotation) of hands and head from packets, but the cursor and the keyboard information are not directly (explicitly) included in the motion data, even though they have certain fixed relations with the motion data (hands or head). 
We could attempt to leverage the visual outputs to observe these hidden attributes.
However, this presents a challenge due to the difficulty in obtaining precise direct observations.
For example, just by looking at the visual outputs, it is hard to gauge the distance between two virtual objects, and an error of just 10 cm can shift a key prediction by three keys (a key has a side length of about 3cm) and render the attack ineffective. 

We present a method to solve this challenge and measure the ``hidden attributes'' with visually verifiable tests.
Specifically, we design geometric tests, which move virtual objects in specific ways to visualize geometric properties, so that we can verify our measurements visually.
We perform the measurements in two steps: (1) Cursor Measurement: we measure the relations between cursor and hand, and uncover how we can convert the hand's \transform that is present in packets to the cursor's \transform. 
(2) Key Measurement: we use the cursor as a reference point to measure the keys. 

\newpar{Cursor Measurement} Cursor measurement is the problem of finding the fixed \textit{offset} (i.e., a constant transformation in position and rotation) between the hand's \transform in packets (referred to as $\transform_{hand}$) to the cursor's \transform (referred to as $\transform_{cursor})$.
From inputs, we can only manipulate the controller's \transform (referred to as $\transform_{controller}$), which has a fixed \textit{offset} to both the hand and the cursor.
These three \transform values can be all different (in position and rotation), as seen in Figure~\ref{fig:3point_problem}.
At a high level, we can solve this problem by first finding the \textit{offset} between the hand and the controller, then finding the \textit{offset} between the controller and cursor, and combining these two \textit{offsets} to get the \textit{offset} between the hand (which appears in the packet data) and the cursor. 
The detailed steps are as follows:

\begin{figure}[h]
\centering
\includegraphics[width=0.33\textwidth]{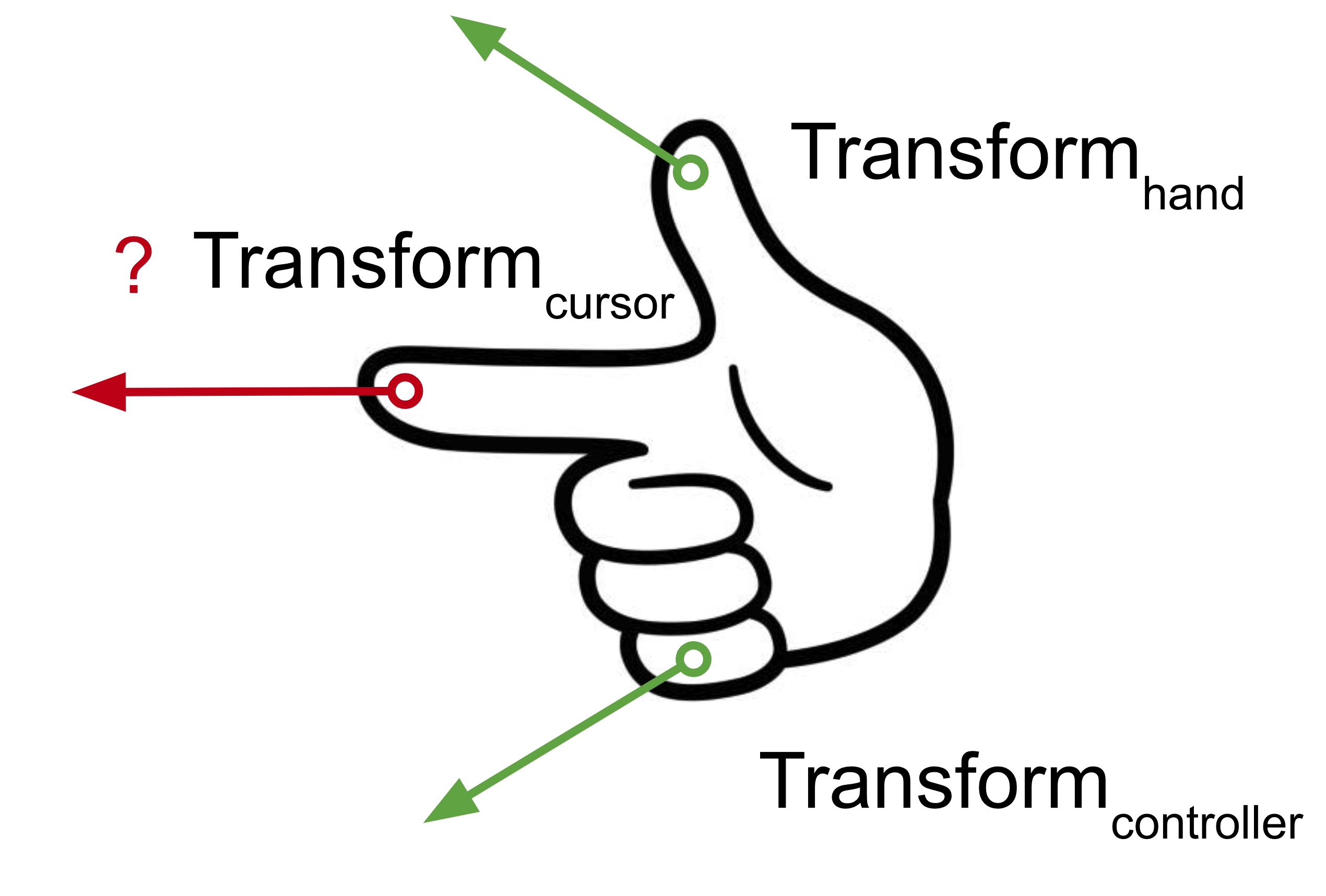}
\caption{We need to uncover conversion from hand's \transform in packets to cursor's \transform by manipulating controller's \transform. }
\label{fig:3point_problem}
\end{figure}


(1) Find the \textit{offset} between $\transform_{hand}$ and $\transform_{controller}$.
Since we can read out $\transform_{hand}$ from packets, it is straightforward to find this \textit{offset} by inputting one sample value of $\transform_{controller}$ and observing the corresponding $\transform_{hand}$, then calculate the constant \textit{offset} in position and rotation between them. 

\begin{figure}[h]
\centering
\includegraphics[width=0.45\textwidth]{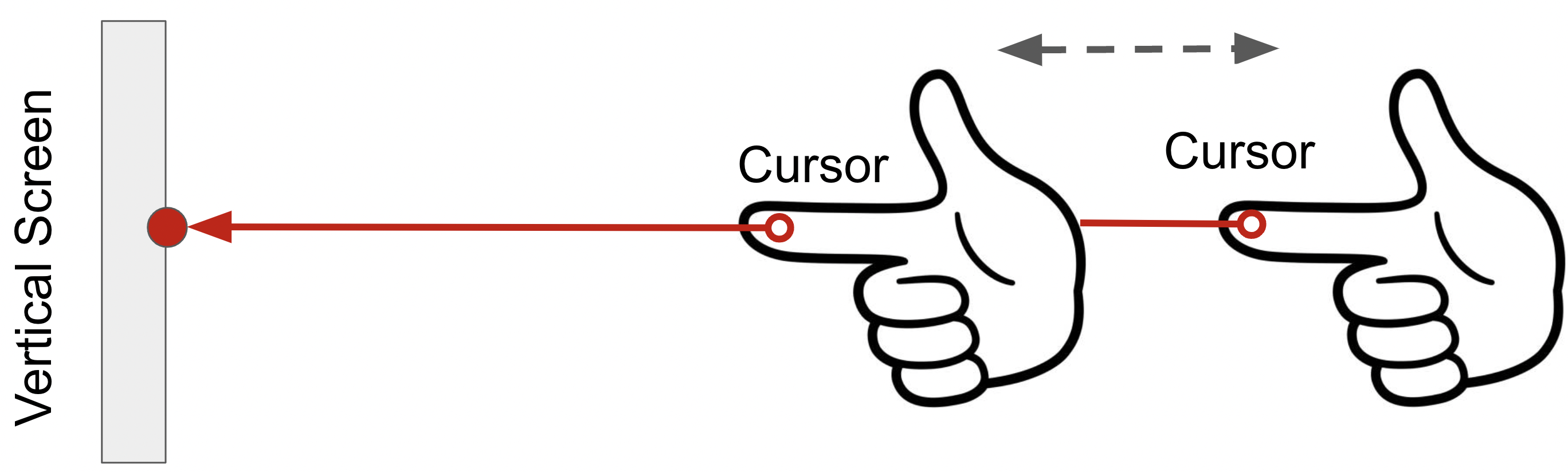}
\caption{Geometric Test 1: test if the guessed cursor orientation is correct by moving the hand along the guess.}
\label{fig:geometric_test1}
\end{figure}

(2) Find the offset between $\transform_{cursor}$ and $\transform_{controller}$. This can be broken into three steps: 
(a) We start with a guess for one part (i.e., position or rotation) of this \textit{offset}. 
For example, if we want to measure the rotational offset, we can start with the guess that the controller and cursor have the same orientation (offset is zero).
(b) Next, we perform a spatial binary search of the real offset.
That is, we plug in the guessed offset as a part of controller inputs and run geometric tests to visually inform us whether our guess is off from the real offset, and in which direction the real offset is.
For instance, we can perform the geometric test as seen in Figure~\ref{fig:geometric_test1} and move the hand along the initial guessed direction. 
If the offset is non-zero, we are not moving the cursor along its pointed direction, and the reticle (cursor's projection on a screen) will move.
(c) Then, we iteratively adjust our guess until the geometric test shows that the guessed offset aligns with the real offset.
Continuing from the example in (b), we can keep adjusting the guess based on which direction the reticle deviates, until the reticle does not move, which is easy to verify visually.
At this point, we know that the hand is moving along the cursor's orientation, and thus our guess is correct.
To learn more about other geometric tests we use, see Appendix~\ref{appendix:measure_cursor}.
 

(3) Calculate our target \textit{offset} by combining the two obtained \textit{offsets}. 
The \textit{offset} in (1) takes us from $\transform_{hand}$ to $\transform_{controller}$, and the \textit{offset} in (2) then takes us from $\transform_{controller}$ to $\transform_{cursor}$, so combining them takes us from $\transform_{hand}$ to $\transform_{cursor}$, which is our goal.


\newpar{Key Measurement} 
Key measurement is the problem of finding the positions of the four corners of a key
at a fixed player location.
The four corners of a key determine the key's location, as they define a bounded plane region as the key's region. 
For this problem, we need to know the cursor's \transform, and use it as a reference.
To measure a corner of a key, we (1) position the cursor at an arbitrary position and find a way to point the cursor at the corner (again through a spatial binary search procedure), then (2) find the distance between the cursor and the key corner, which can be done by using a spatial binary search with another geometric test (see Appendix~\ref{appendix:measure_keyboard}), and lastly (3) calculate the key's position by adding a vector that is along the cursor's pointing direction and of the length equal to the distance to the cursor's position.


\newpar{Keyboard Measurement} Keyboard Measurement is the problem of finding each key's position when the player is in an arbitrary position and orientation.
This problem is straightforward to solve once we are able to perform \textbf{Key Measurement}. 
We can use the originally measured keys as a baseline, and repeat measuring all keys' positions at multiple player positions and rotations.
This allows us to calculate how the changes in player positions and rotations transform into changes in the keys' positions.
Therefore, for a new player's position and rotation, we simply calculate the changes in position and rotation from the baseline, and apply this transformation.

\section{Data Collection}
\label{experiment}
We performed data collection experiments to assess the effectiveness of our attack in real-world scenarios.
Specifically, we conducted a user study to collect real-world typing activity data from Rec Room, one of the most popular multi-user VR applications, as the main experiment. 
We have also conducted further experiments to demonstrate that our attack works well in three additional applications of various genres and in realistic scenarios.

\newpar{Data Collection Setup}
During the user study, we installed and ran Rec Room in VR mode through SteamVR on a Windows PC, with an Oculus Quest 2 connected via Quest Link (we will refer to this as the \textit{Victim PC} later). 
We installed and ran Rec Room in non-VR mode on another Windows PC to simulate the attacker behavior (we refer to this as the \textit{Attacker PC}). 
The two PCs then join one private virtual room, as mentioned in our threat model.
After we begin capturing network traffic with Wireshark on the \textit{Attacker PC}, the participant, wearing the Quest 2 headset, starts entering the prepared text into Rec Room's chat functionality using the virtual keyboard.
Lastly, to serve as a reference during the data labeling process, on the \textit{Victim PC}, we use VCR to record the VR tracking data during each session.


\newpar{User Study Recruitment}
We conducted a user study to collect VR typing data from volunteers using the \textit{Victim PC}. 
This study has been approved by our university's Institutional Review Board (IRB). 
Through email advertisements via our institution's email list, we recruited 20 participants of varying ages, heights, genders, and experience with VR from our institution's campus. 
Prior to the experiment, participants were provided with instructions on how to operate the Quest~2 headset and controllers. 
Additionally, they were guided on the typing functionalities and practiced typing within Rec Room.
Before the start of the experiment, the participants were also informed that their typing-related information would be recorded and that they could stop the study at any time. 
To minimize bias and more closely mirror real-world typing scenarios, participants were not initially informed of the study's exact purpose. 
Instead, they were simply told that the goal of the study was to examine VR typing behavior.
After the participants completed all the typing experiments, we debriefed them about the real purposes of the study and asked them to fill out a survey to understand their perceptions of our attack.
The study lasted approximately 60 minutes per participant, and each participant who took part in our study was compensated with a \$20 gift card. 



\newpar{User Study Typing Trials}
For each experiment, a participant started by opening the Rec Room chats to perform typing trials.
In each trial, the participant was presented with a prompt and was asked to type all the characters in the prompt into the chat.
In total, each participant completed 65 trials, of which 30 trials were with number prompts, 20 were with password prompts, and 15 were with sentence prompts.
To simplify the setup, all prompts were shown to the participants in the same chat in which they performed the trials, so they could see the prompt while typing. 
We had a researcher continuously observing the participants' inputs and asking the participants to retype a prompt if they mistyped it (e.g., they typed ``124'' when the prompt was ``123''). 
By doing this correcting step, the prompts could conveniently serve as the ground truth without modifications during the evaluation of the attack accuracy later. 


As previously mentioned, the participants engaged in trials with three types of prompts, each representing a common typing behavior:


(1) \textit{Numbers}: The prompts are random numbers with lengths of 3 (e.g., ``823''), 9 (e.g., ``804458083''), and 12 (e.g., ``595397360820'') digits, with 10 prompts per length.
    This simulates entering credit card information. 

(2) \textit{Passwords}: The prompts are passwords with characters, numbers, and punctuation marks. 
    To emulate real-world password typing activity, we generated 10-17 characters long combinations of English words with numbers and punctuation marks, using the Memorable Password Generator~\cite{memorable_password_generator}.

(3) \textit{Sentences}: The prompts are English sentences that consist of 3 words (e.g., ``Clouds are drifting''), 6 words (e.g., ``The sunset today looked absolutely stunning''), and 9 words (e.g., ``My cat just did the funniest thing this morning''), with 5 prompts for each length.
    These sentences are generated by ChatGPT~\cite{openai2022chatgpt}. 
    This experiment is to simulate the scenario of typing in private chats.

After the study concluded, we asked the participants if they wished to exclude their data and address any questions they might had. 
At the time of submitting this paper, we have not received any requests for excluding data.


\newpar{Data Labeling}
During the study, we utilized a timing application to mark the beginning and end of each trial. 
This ensured only data from the trials was considered for labeling.
Then, we check the recording of tracking data during the trials to verify that the number of clicks corresponds to the expected number of clicks from the prompts. 
For example, if the prompt asks the user to type ``123,'' then we should expect three clicks in the recording of tracking data. 
If the number of clicks matches, we then label each click using the characters in the prompts along with its timestamp in the tracking data recording, so that we can later compare the clicks detected in the network traffic with the ground truth clicks.
However, if the number of clicks does not match, it signals that the prompt was mistyped, which may happen when the prompts are long, and the researchers did not spot the errors in what the participants typed during the study.
The data for such trials is discarded and not considered for evaluation. 
However, these cases are rare.
In total, we dropped a combined 638 clicks out of 22,730 clicks (29 out of 1,300 trials) collected from the 20 participants in our study. This leaves 22,092 keystrokes (from 1,271 trials) for the evaluation.

\newpar{Other Practical Scenarios}
We also conducted an experiment to demonstrate the practicality of our attack in scenarios where (1) there are multiple users in a virtual room, and (2) the attacker does not see the victim users.
We created the same experiment setup as the \textbf{User Study Typing Trials} experiment 
but with two modifications: (1) we placed another four different victim users in the room instead of one; 
(2) the attacker faces a wall, thus they are unable to see any other users. 
We then conducted a user study with the same procedure as the main experiment but with two participants typing concurrently (the other two users in the room were dummy users).
Since the attack worked exactly the same in this setting, we ran this experiment only once. 
In total, we collected 2,431 clicks from the participants.

\newpar{Other Multi-user VR Applications}
To show that our attack generalizes beyond Rec Room, we tested it on three additional popular applications from diverse genres: Galaxity~\cite{galaxity}, Sing Together: VR Karaoke~\cite{sing_together}, and oVRshot~\cite{ovrshot} (see Appendix~\ref{appendix:additional_app} for more details).
Specifically, we conducted an end-to-end attack on each of these applications following the same attack procedure (Section~\ref{sec:approach}) and setup of the experiment on Rec Room (the \textbf{User Study Typing Trials} experiment).
Among these applications, we targeted keystrokes on the full keyboards of two (Galaxity and Sing Together: VR Karaoke), whereas in oVRshot, attention was directed to the numeric keyboard, given that it is the only type available.
During the typing experiment, which was conducted with three participants for each application, 
we collected a total of 3,423 clicks for Galaxity, 3,473 clicks for Sing Together: VR Karaoke, and 760 clicks for oVRshot (numbers trials only).
Note that for passwords and sentences trials, we selected random words that can have slightly different lengths, which leads to a slight variance in total keystrokes for different applications.

\label{sec:steathy_robust}




\section{Evaluation}
In this section, we evaluate our attack and analyze various factors that may affect the attack accuracy. 
We have analyzed whether the attack accuracy is affected by packet drop rates, key position on the keyboard, and typing speed.

To evaluate the attack accuracy, we use top-$k$ accuracy, which assesses how many of the user-typed keys are correctly predicted within our attack's top-$k$ predicted keys, sorted by distance from the position of the cursor's projection on the keyboard. 
In other words, the top-$k$ accuracy is calculated as (number of successfully inferred keys) / (number of total keys).
If the click is not identified by the keylogging attack, we automatically mark the prediction as wrong. 


\begin{table}[ht]
\begin{center}
\begin{tabular}{llll}
\toprule
          & Top 1   & Top 3   & Top 5   \\
\midrule
Numbers   & 96.78\% & 97.98\% & 98.23\% \\
Passwords & 97.42\% & 97.88\% & 98.17\% \\
Sentences & 98.16\% & 98.41\% & 98.52\% \\
\midrule
Total     & 97.62\% & 98.15\% & 98.34\%\\

\bottomrule
\end{tabular}
\caption{Keylogging attack accuracy from 20 participants across three different typing tasks.}
\label{main_result}
\end{center}
\end{table}

Table~\ref{main_result} presents the overall accuracy of our attack based on 22,092 keystrokes collected from all 20 participants in the user study. 
The evaluation setup is consistent with the character-level evaluation in prior research on VR keylogging attacks~\cite{luo2022holologger,gopal2023hidden,wu2023privacy,slocum2023going}.

Across all three types of prompts, our keylogging attack consistently exhibited a very high accuracy, inferring 97.62\% (21,567 out of 22,092 keystrokes) of all keystrokes correctly with top-1 prediction, 98.15\% (21,683 out of 22,092 keystrokes) with top-3 prediction, and 98.34\% (21,726 out of 22,092 keystrokes) with top-5 prediction. 
Against a 2.13\% random guess baseline for one-key inference, our attack's effectiveness increases over 45 times.

Although our attack is nearly perfect in accuracy, there are a few incorrect predictions.
This can happen for one of the following two reasons: 
(1) Loss of motion data precision due to lossy compression of the data: To ensure stable network performance, \vrsocialapp such as Rec Room compress motion data items before transmitting them over the Internet. 
When this data is decompressed by the adversary client, it will be slightly off compared to the ground truth motion.
This slight imprecision in motions and cursor positions might cause inaccurate predictions.
(2) Stale hand motion update: From our experiment, we observed that some hand motions are not updated (propagated) immediately. 
This might happened because of the lag on the victim client or the interpolation mechanism applied on the previous motion update.
Consequently, when this happens, the hand's \transform associated with a click tends to be closer to the previous click rather than the intended target click and leads to incorrect keystroke predictions by our attack. 

It should be noted that the loss of motion data precision due to the compressed data is not significant.
Additionally, the cases of stale hand motion occur with a low probability. 
This explains why only 2.38\% of the keystrokes in our experiment are predicted incorrectly.

These findings from our attack highlight the significant risk of privacy breaches due to motions detected in the network packets, emphasizing the urgent need for protection against such vulnerabilities.

\subsection{Our Attack is Robust Under Traffic Loss}
\label{sec:traffic}

\begin{figure}[h]
\centering
\includegraphics[width=0.40\textwidth]{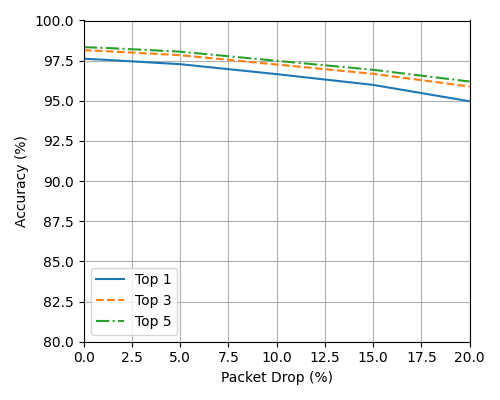}
\caption{\label{accuracy_vs_drop}Our attack is robust against random packet drop, achieving a top-1 accuracy of 94.97 percent even when 20 percent of the packets are dropped.}
\label{fig:drop}
\end{figure}

Since the attack is conducted remotely, it is crucial for us to study the packet loss during the transport of motion data and how it affects the accuracy of our attack. 
According to a recent article by Obkio~\cite{obkio}, apps with a packet loss rate of more than 10\% are considered unusable~\cite{obkio}. 
To stress-test our attack, we conduct an analysis by dropping random packets at an even higher rate of up to $20\%$ from the participants' network captures.
The corresponding attack accuracy is illustrated in Figure~\ref{fig:drop}. 
From the result, we observe that the attack accuracy after 20\% of random packet drops rate still remains very high (94.97\% top-1 accuracy), showing that our attack is robust even under significant packet loss.

In total, we lost 2.65\% of the keystroke recovery accuracy due to the following two reasons: 
(1) Loss of brief keystroke data due to packet loss: In instances where the click duration is extremely short, the motion data corresponding to this key might only be present in a few packets. 
If there is packet loss, these packets might be dropped. 
As these packets are never received, they cannot be used for key recovery; (2) Packets after a click motion are considered as the actual click: 
By definition, a click occurs the moment the victim presses the key beyond the click threshold, as mentioned in Section~\ref{click_mechanism}. 
However, a packet loss might occur at the moment of a click.
As a result, the initial packet representing the click might be missing.
In this case, our attack identifies a subsequent packet during the click period as the click packet (i.e., the attack thinks this later packet is the first frame of a click). 
However, the subsequent packet might show a minor difference in the hand's \transform from the original click packet, resulting in a variation in the click motions and causing a wrong keystroke identification.

It should also be noted that we observed only an additional 0.42\% of clicks undetected when a 20\% packet loss is introduced. 
Also, the situation where selecting later packets results in a decrease in attack accuracy only arises in specific cases. 
It happens only when the click motion changes rapidly and the initial few packets are lost. 
This is why our attack accuracy is only slightly affected by the significant packet drop rate. 

\subsection{Keyboard Layout Affects Attack Accuracy}
\label{sec:keyboard}

Given that VR users utilize their hands to control the cursor to select keys on the virtual keyboard, we hypothesize that keys positioned farther away from the user may be more challenging to predict accurately as these keys are relatively smaller in the user's field of view. 
Our intuition is based on the fact that the cursor, when pointing at a distant key, adopts a more tilted angle.
Consequently, this might increase the likelihood of an incorrect prediction by the attack as the inference to the relatively smaller keys is more sensitive to deviations introduced by the transmission process.

\begin{figure}[h]
\centering
\includegraphics[width=0.4\textwidth]{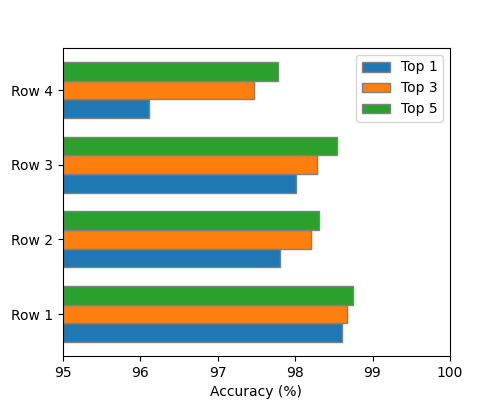}
\caption{\label{accuracy_vs_row}The keys positioned further away from the user are relatively more difficult to infer compared to the closer keys.}
\label{fig:row}
\end{figure}


To test this hypothesis, we categorized the keys on the keyboard into four rows based on their layout. 
Row~1 starts with keys `z,' `x,' and `c,' Row~2 starts with keys `a,' `s,' and `d,' Row~3 starts with keys `q,' `w,' and `e,' and Row~4 starts with keys `1,' `2,' and `3.' 
The accuracy rates for each row are displayed in Figure~\ref{fig:row}. 
Although the attack accuracy remains high (with a top-$1$ accuracy of~~>96\%) across all rows, there is a noticeable trend: the accuracy diminishes as keys are positioned further away from the user (for instance, top-$1$ accuracy drops from 98.6\% in Row 1 to 96.12\% in Row 4).

\subsection{Our Attack Remains Robust Under Varying Typing Speed}
\label{sec:user_behaviors}

We also studied how the typing speed impacts the accuracy of our attack. 
Intuitively, predicting faster clicks might be more challenging since the cursor moves more quickly. 
Even a slight lag could significantly affect the position of the cursor's projection on the keyboard.
We present accuracy data grouped by click duration percentiles (e.g., the 0-20 percentile group comprises the fastest 20\% of clicks). 
Contrary to our initial expectations, we did not observe a significant effect of typing speed on the attack accuracy.
This finding could be attributed to the fact that even the fastest click (with a duration between 0.255 and 0.721 seconds for the 0-20 percentile group) is captured in multiple motion updates, given that motion updates occur frequently (15-20 updates per second, or 0.05 to 0.067 seconds between updates).

\begin{figure}[h]
\centering
\includegraphics[width=0.4\textwidth]{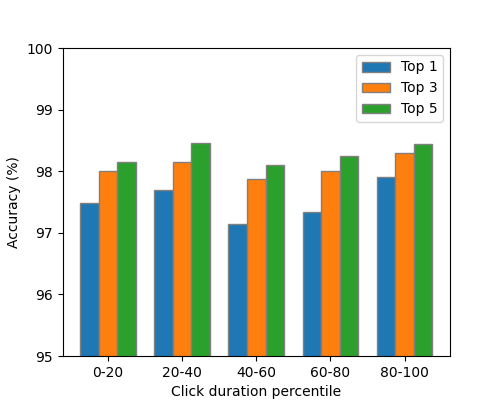}
\caption{\label{accuracy_vs_speed}Our attack is unaffected by typing speed, with similar accuracy across different click duration percentiles.}
\label{fig:speed}
\end{figure}


\subsection{Our Attack Can Be Stealthy Under Practical Scenarios}
For the experiment setting in which (1) there are multiple users typing concurrently in the same room and (2) the attacker is facing the wall, the attack has comparable performance to the \textbf{User Study Typing Trials} experiment.
Our attack correctly inferred 97.53\% (2,371 out of 2,431) of all keystrokes with top-1 prediction, 99.51\% (2,419 out of 2,431) with top-3 prediction, and 99.59\% (2,421 out of 2,431) with top-5 prediction.
This result shows that (1) our attack is not affected by the number of participants in the room, and (2) even when the attacker cannot see the victim, the motion data is still received and can be used to recover the keystrokes. 
Thus, we conclude that our attack is applicable in a multi-user environment, and the attack is stealthy since the attacker can hide anywhere in the room while performing the attack.

\subsection{Our Attack Generalizes Across Applications}

\begin{table}[ht]
\begin{center}
\begin{tabular}{llll}
\toprule
          & Top 1   & Top 3   & Top 5   \\
\midrule
Galaxity   & 98.25\% & 99.71\% & 99.73\% \\
Sing Together: VR Karaoke & 98.27\% & 99.97\% & 99.97\% \\
oVRshot & 99.07\% & 99.61\% & 99.61\% \\

\bottomrule
\end{tabular}
\caption{Keylogging attack accuracy from three participants typing in the selected three additional applications.}
\label{multiapp_result}
\end{center}
\end{table}

By performing our proposed attack on Galaxity~\cite{galaxity}, Sing Together: VR Karaoke~\cite{sing_together}, and oVRshot~\cite{ovrshot}, we have successfully recovered the motion data from the collected traffic and infer the key participants typed. 
For more details on the recovered packets, please refer to the examples in Appendix~\ref{appendix:example_photon_packet}.

The performance of our attack is as detailed in Table~\ref{multiapp_result}.
The slightly varied attack accuracy across applications may be attributed to differences in keyboard layouts and participant typing habits.
Nevertheless, all attack results demonstrated performance comparable to our main experiment on Rec Room, underscoring the generalizability of our attack across different applications.

\section{Keylogging is Possible Even with Partial Reverse Engineered Packets}
\label{machine_learning}

In the previous sections, we demonstrated the effectiveness of performing a keylogging attack by fully reconstructing the typing process. 
However, the involved steps require some manual effort and can be time-consuming. 
If an adversary aims to execute our remote keylogging attack rapidly, with minimal manual reversing effort, the keylogging attack can still be performed with the help of machine learning.
This approach operates under an additional assumption compared to the threat model introduced in Section~\ref{threatmodel}. 
In particular, we allow the adversary to obtain some motion data from a victim's typing that is paired with the actual text being typed, serving as labeled (ground truth) data.
This assumption is commonly accepted and used in previous VR keylogging work~\cite{slocum2023going, al2021vr, zhang2023s}. 
In the case of \vrsocialapp, adversaries can collect the necessary data by chatting with the victim (via the application chat) and capturing any messages with the corresponding motion data. 
Since the victim can perform multiple tasks without closing the menus, and the keyboard position is fixed once they open the menu, this collection step allows the adversary to use the labeled data to predict other keystrokes on the same keyboard location.

\begin{figure}[h]
\centering
\includegraphics[width=0.45\textwidth]{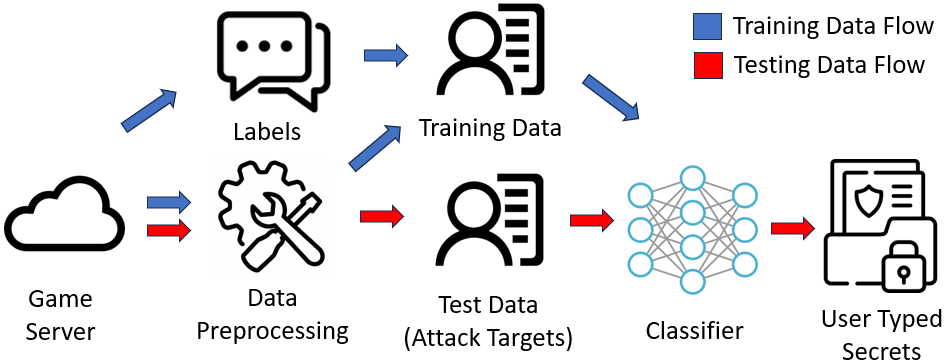}
\caption{Machine Learning Attack Overview: The blue arrows illustrate the model training process, while the red arrows describe how an adversary might utilize the trained model to infer user keystrokes.}
\label{fig:mlatt}
\end{figure}

\newpar{Attack Setup}
To show that our attack works with limited reversing, we use the example of Rec Room to further demonstrate this attack.
Figure~\ref{fig:mlatt} outlines the process. 

To build the training dataset for the machine learning model, we first collect raw packets and their corresponding labels (the typed text) using the aforementioned phishing approach. 
To preprocess the raw packets, we follow the steps from Section~\ref{sec:approach} that do not require significant manual effort and can be consistently applied across applications:

(1) Filter for the packets that contain motion data, as explained in Section~\ref{subsec:packet}. 

(2) Use the protocol parser for Photon, as referenced in Section~\ref{subsec:field_extraction}, to parse the packets, given that Rec Room employs the Photon library for motion updates. 
This step parses the packets into different fields of readable formats (e.g., integers, floats), except for the custom object fields, which are left as raw bytes.
This step is easy because the Photon protocol parser is a readily available tool that can be applied to all applications that use the Photon libraries.

(3) Run the \textbf{Precise Input Control} experiment (see Section~\ref{precise_input_control}) to identify the fields that correspond to the click data (which are stored in floats), and the custom object fields that contain the motion data.
This step is easy because the experiment can be run automatically, and we only need to manually observe the associations between the input data and the fields.

(4) Determine the timing of clicks by performing click detection, which is done in the same way as in Section~\ref{subsec:clickpos}.
We can use this information to find the packets that correspond to clicks and select the custom object field that contains the hand motion in these click packets (e.g., select the custom object field that contains the right-hand motion if the participant uses the right hand to click). 
Recall that click identification is straightforward via standard Unity APIs (see Section~\ref{click_mechanism}). 

Compared to the attack described in Section~\ref{sec:approach}, the resulting data will no longer have the following information:
(1) The parsed motion data in a readable format.
This requires reversing the parsing of custom objects, and this process needs to be redone for each application, as mentioned in Section~\ref{subsec:field_extraction};
(2) The \transform of the cursor and key.  
This is because we do not have motion data in a ``meaningful'' format, so we cannot perform calculations on top of it (e.g., measure cursor and keyboard offsets).

After these data preprocessing steps, for each click, the resulting data provides only a custom object field in raw byte format that contains the motion data associated with typing.
The resulting data is paired with the corresponding labels to create a dataset to train a machine learning classifier.
Subsequently, we gather more traffic as the victim types. 
These attack targets (traffic with keystrokes) can be processed in the same manner as the training data and fed into the trained machine learning classifiers to infer corresponding keystrokes.

\newpar{Attack Evaluation and Results}
For evaluation purposes, we utilize the traffic traces collected from the user study and divide them into an 8:1:1 ratio for training, testing, and validation. 
Then, we follow the attack steps above to carry out the attack and evaluate the attack result using the top-$k$ accuracy of the testing data.

 To identify the best machine learning model for the task, we compare the attack results from various models, including SVM~\cite{cortes1995support}, GBM~\cite{ke2017lightgbm}, MLP~\cite{rosenblatt1958perceptron}, and CNN~\cite{lecun1998gradient}, to the Random Guess baseline. These models are trained with a learning rate of $10^{-4}$ for 500 epochs, from which we chose the checkpoint with the highest validation accuracy as the best attack model.



From the results in Table~\ref{ml_result_all}, we observe that all ML models extract the victim-typed keystrokes significantly better than the random guess baseline. 
Additionally, we observe that the CNN model performs the best, with a 68.07\% top-1 accuracy, 85.96\% top-3 accuracy, and 90.28\% top-5 accuracy in predicting keystrokes. 
This result highlights that machine learning models like CNN can effectively learn about mapping the bits in packets to corresponding motions and click positions, so the model can predict keystrokes with reasonable accuracy. 

Our CNN model structure consists of three convolutional layers. Each layer has a one-dimensional kernel size of three, with 32, 64, and 128 neurons in each layer.
Subsequent to these convolutional layers, the architecture includes two fully-connected layers and a softmax layer.
The output space of this CNN model corresponds to the number of classes, equivalent to the distinct keys on the keyboard (47 in total). 

The success of CNN might be attributed to its ability to learn from small datasets with generalizability~\cite{alzubaidi2021review}. 
Also, the CNN model might be able to learn from the features with significant effects on the key presses (e.g., those bits directly related to the \transform of the typing hand)~\cite{alzubaidi2021review}, which can be effective in capturing relevant parts from the partially reversed data items that contain non-parsed custom fields used in this task.
\begin{table}[ht]
\begin{center}
\begin{tabular}{llll}
\toprule
          & Top 1   & Top 3   & Top 5   \\
\midrule
Random Guess   & 2.13\% & 6.38\% & 10.64\% \\
SVM & 44.87\% & 64.47\% & 71.57\% \\
LightGBM & 46.49\% & 66.24\% & 71.61\%\\
MLP & 61.99\% & 79.81\% & 85.34\% \\
CNN     & 68.07\% & 85.96\% & 90.28\%\\
\bottomrule
\end{tabular}
\caption{Comparison of machine learning models on inferring the keystrokes with partially reversed data in raw bytes.}
\label{ml_result_all}
\end{center}
\end{table}

The accuracy is still lower compared to the full extraction keylogging attack we proposed earlier, which uses high-fidelity typing-related motion data. 
This is due to two reasons: 
(1) From the prior experiment, we find that a slight deviation in the motion data received by the adversary can cause an incorrect prediction since the motion data of typing nearby keys is similar due to the relatively small space of keys. 
As the model is trained on partially reversed data, the data might have less learning signal to predict the key and might introduce errors. 
(2) As the model is trained on data items collected through chatting with a victim, the quantity of the data is also limited.
This can potentially lead to the overfitting of the model and cause a deviation when inferring the motion data that the model has never seen, leading to a decrease in prediction accuracy. 
However, even with the drop in accuracy, the attack can still very effectively infer the victim's keystroke. 


We have also performed further analysis of the attack results in Appendix~\ref{mlother}, focusing on inferring keystrokes with different amounts of training data, different hands, and during different typing tasks.

\section{Discussion}
\label{sec:discussion}
\newpar{User Awareness and Concerns} 
From the results of the survey (see Appendix~\ref{survey} for the survey questions), 
none of the participants had ever imagined that their input in \vrsocialapp could be stolen by adversaries. 
However, after disclosure, 12 out of 20 participants expressed concerns regarding this type of attack (Score > 4).
These findings suggest that VR users do recognize the potential harm from such attacks. 
The user study results highlight the importance of educating VR users about potential security and privacy threats, such as our keylogging attacks in \vrsocialapp. This education can serve as a warning, encouraging users to exercise caution and prevent privacy leakage.

Privacy policies of VR applications play a crucial role in educating users about their data privacy. 
However, to the best of our knowledge, major VR applications often fail to clearly explain sensitive data like motion data in their privacy policy.
Furthermore, Trimananda et al.\cite{trimananda2022ovrseen} find that approximately 70 percent of PII data sent in traffic of VR applications are not properly disclosed. 
This highlights the need for better disclosure and compliance efforts from VR developers.

\newpar{Impacts of Our Attack} 
We have reported our attack findings to all of the applications we evaluated, as well as SteamVR and Unity. 
Rec Room, Sing Together: VR Karaoke and SteamVR responded, acknowledging the vulnerability.
Furthermore, Rec Room classified it as \textit{\uline{P3 severity (Medium: Vulnerabilities that affect multiple users and require little or no user interaction to trigger)}}. 
This underscores the fact that our attack is feasible and poses a privacy risk to numerous users.
At the time of writing, we still have ongoing discussions with SteamVR and Unity.

It should also be noted that our reverse engineering approach is not limited to recovering typing-related motion; 
it can also recover other types of sensor data shared over the internet, such as voice data. 
Our tool can serve as a baseline for future studies aiming to further reveal the potential threats posed by the misuse of sensor data in VR.

Moreover, our attack poses a threat not only to users in applications which we have performed our attack. 
For multi-user VR applications, the function of correctly synchronizing avatar movements and displaying them to other users is foundational, serving as the cornerstone of real-time interaction within the application. 
To accurately render user avatars and their movements, it's important to note that even if motion data is encrypted when sent over the network, it will eventually be decrypted on every client.
Therefore, attackers can recover motion data (of other users) received by their own client, which allows them to fully track how a victim's avatar moves when selecting a key.
Since our attack targets and exploits the transmission of motion data, which is fundamental to the design of multi-user VR applications, the threat posed by the attack is not limited to the applications we have evaluated.
In addition to them, we have analyzed the functionality of \textbf{30} multi-user VR applications from various platforms, including Oculus and Steam. 
Of the applications we studied, \textbf{18} offer virtual keyboard typing functionality and transmit motion data online, making their users potential victims of our attack. 
Additionally, the wide support of overlay system messaging apps like Steam Chat or Messenger on VR devices makes all \vrsocialapp potentially susceptible to our attacks, since the applications may transmit motion data when these overlays are activated.

\newpar{Our Attack is Applicable Regardless of Physical Setup} 
It is important to note that the transmitted motion data represents the movements of in-game objects (e.g., avatar hands), rather than the physical movements of devices (e.g., controllers).
As such, the motion data remains consistent irrespective of the attacker's or victim's physical setup (e.g., VR device, WiFi router), 
with the setup only affecting the data's quality (e.g., refresh rate).
In Section~\ref{sec:traffic}, we have shown that our attack is robust against degraded motion data quality.


\newpar{Defenses}
To the best of our knowledge, no current defense mechanism directly addresses our attack, and common data protection defenses struggle to mitigate our attack for the following reasons:

(1) Encrypting network traffic is not enough: as discussed by Trimananda et al.\cite{trimananda2022ovrseen}, if an adversary controls the client application, they can intercept packets during the key exchange, thereby obtaining the encryption key. 
This enables the adversary to decrypt subsequent traffic to their client and access motion data, facilitating the described attack.
Interestingly, among the applications we examined, only VRChat encrypts the user's motion data within the packets. 
This highlights a general oversight by \vrsocialapp with respect to the potential privacy breaches related to motion data.

(2) Differential privacy comes with utility trade-offs: by introducing noise to all motions, differential privacy can decrease attack accuracy and mitigate potential motion-related privacy breaches.
However, as highlighted by Nair et al.\cite{nair2022going}, incorporating differential privacy to protect the motion data can lead to utility drops.
That is, it can lead to altered and inaccurate avatar movements, and can potentially reduce the quality of the immersive experience in VR applications.

We propose a defense to mitigate this attack: full blockage of motion updates during typing activities. 
That is, a user's motion data should not be sent to remote clients when the user is typing, and perhaps an idle or random animation can be sent instead. 
While this solution may sound trivial, it requires efforts from developers, game engines, and VR systems alike. 

For typing activities using an application's built-in typing functionalities, the burden of defense is now put on every developer to identify scenarios in which typing happens and fix them, which can be error-prone. 
Therefore, it may also be necessary for game engines to provide a standard API and defense mechanism for the typing functionality. 

However, as previously mentioned, just protecting motion data for an application's built-in typing functionalities is insufficient because of the wide support of system overlays like Steam Overlay~\cite{steam_overlay}. 
As the application is still active, motion data is still updated when the overlay is launched, and typing activities inside the overlay are also vulnerable to our attack.
In this case, the application cannot know whether the users are typing and when to block motion updates.
Consequently, a standardized system-level API for typing detection also needs to enable communication between the system and applications.

\newpar{Adoption of Defense}
As of August 2024, SteamVR and Rec Room have created patches to defend against our proposed attack, following our suggestions of the aforementioned defense. 
Specifically, in SteamVR Beta 2.7.1~\cite{steamvr_beta_271}, SteamVR restricts VR applications from accessing hand motion data when the Steam keyboard is opened in Steam Overlay, so that the applications cannot inadvertently synchronize the typing motion data with remote users.
Similarly, in Rec Room's ``My Little Template'' Edition update~\cite{recroom_defense_update}, Rec Room stops synchronizing a user's hand motion data with remote users when the user is typing in a sensitive field (e.g., password). 



\newpar{Limitation and Future Work}
At this stage, our method predominantly targets a common input approach: using controllers to select inputs from virtual keyboards, which highlights vulnerabilities in motion data transmission. 
While alternative input methods exist, such as using one's hand to tap keys on a virtual keyboard, these mechanisms still rely on motion data for input. 
Therefore, the foundational concept of our attack remains valid. 
Thus, our technique should also be effective in extracting typed secrets from these variations.

Also, our work primarily showcases that keylogging attacks can be performed with motion data extracted from network traffic. 
However, the use of motion data is not limited to updating remote avatars; it also plays a role in various stages of a VR system, such as rendering, haptic feedback, and video recording.
We will leave this as future work to further explore other remote channels available for keylogging attacks. 


As highlighted by Nair et al.~\cite{nair2023truth}, motion data can lead to other forms of privacy breaches, such as user identification, anthropometric measurements (e.g., height and wingspan), and even demographic details (e,g., age and gender).
It would be valuable to conduct a separate study in the future, exploring how to remotely exploit these vulnerabilities to deduce such information from VR players.
Moreover, we primarily focused on utilizing motion data transmission as a side channel to perform a keylogging attack.

\newpar{Ethics Statement}
Our paper demonstrates a practical keylogging attack that utilizes an inherent issue of VR multiplayer applications.
We have followed a responsible disclosure process to notify affected parties and to mitigate potential harm. 
Also, we believe that our work serves as an important warning to developers and we hope that it will inspire the design of better defenses and secure future applications.
\section{Related Work}
\newpar{Keylogging and Side-channel Attacks on VR/AR Devices}
Given that VR and AR applications continuously gather user motion data, and typing is dependent on hand or controller movements, recent studies have identified potential side channels for keylogging on VR devices, such as using malware running in the background on the victim headset to collect a user's hand or head movement for keylogging~\cite{luo2022holologger,slocum2023going,wu2023privacy} or using system-side channels, like rendering performance counters~\cite{zhang2023s} or channel states~\cite{al2021vr}.

It is also worth noting that methods of keylogging through video capture of user typing have been extensively researched in previous studies~\cite{gopal2023hidden, meteriz2022keylogging, ling2019know}. 
However, previous studies operate under the assumption that adversaries can access a user's local information during typing. 
This may not be applicable in many VR usage scenarios. 
In contrast, our keylogging attack is executed remotely without necessitating any modifications that could alert or impact the victim, rendering our method both practical and discreet. 
Although our attack is performed under a more challenging  threat model, it still achieves comparable performance to the state-of-the-art keylogging attack on VR motion data~\cite{wu2023privacy}.

\newpar{Keylogging and Side-channel Attacks on Other Devices}
Keylogging has also been well-studied on smart devices, including smartwatches, mobile phones, and tablets. 
The techniques for keylogging on these devices also focus on exploiting side channels, such as sensor leakages~\cite{xu2012taplogger,cai2011touchlogger, wang2015mole, maiti2016smartwatch, liu2015good}, video recordings of human typing~\cite{yang2023towards, sun2016visible}, voice~\cite{halevi2015keyboard, schlegel2011soundcomber}, Wi-fi signals~\cite{ali2015keystroke, shen2021wipass}, CPU-related side channels~\cite{paccagnella2021lord, zhang2016return} and GPU-related side channels~\cite{ladakis2013you,zhan2022graphics}. 
Again, these prior efforts are also based on the assumption that adversaries can access and exploit side channels on user local devices. 
This is primarily because the motion data associated with typing does not necessarily need to be exported or shared with another person or client over the Internet. 
In contrast, our attack exploits the unique property of immersive VR environments, which requires broad sharing of motion data for rendering. 
The effect of remote motion data leakage proposed in our work is understudied and unique to \vrsocialapp.

\newpar{Other VR/AR Security or Privacy Issue}
Other than keylogging attacks, security and privacy problems in VR have also received attention recently. 
Regarding privacy in VR/AR applications, Nair et al.~\cite{nair2022going,nair2023unique,nair2023truth} have demonstrated different kinds of privacy threats related to the data-collecting process in VR/AR systems and proposed utilizing differential privacy tools for protection. 
Trimananda et al.\cite{trimananda2022ovrseen} investigated VR app network flows for privacy breaches, while Farrukh et al.\cite{farrukh2023locin} utilized VR/AR spatial maps to identify sensitive environmental details.
Regarding security, studies have centered on Perceptual Manipulation Attacks (PMA)\cite{cheng2023exploring,casey2019immersive}, clickjacking\cite{su2022perception}, and ad fraud~\cite{lee2021adcube}.
Similar to these prior efforts, our keylogging attack utilized unique designs in VR/AR systems. 
However, our method is specifically designed to execute a remote keylogging attack in \vrsocialapp to extract user typing information.

\section{Conclusion}
In this work, we present the first VR remote keylogging attack that targets \vrsocialapp. 
Specifically, our attack can accurately infer user keystrokes by recovering motion data from the attacker's client, {\it without} the need to compromise the victim's device or physically approach the victim. 
In a user study with 20 participants, our attack showcases that, even with rate limitations and the potential for packet loss under a remote attack setting, we can still infer the victim's keystrokes with a nearly perfect top-1 accuracy of 97.62\% and top-5 accuracy of 98.34\%. 
Furthermore, our results indicate that even with minimal manual reversing efforts, an adversary can swiftly deploy this keylogging attack across various applications with an added phishing step using machine learning, still achieving a reasonable top-1 accuracy of 68.07\% and top-5 accuracy of 90.28\%.
We hope this work can assist the VR research community and industry by highlighting potential threats from motion leakage in VR and encouraging the development of more effective defense mechanisms.

\section*{Acknowledgments}

We sincerely thank the reviewers and shepherd for their valuable feedback on the paper.
This work is supported in part by the National Science
Foundation (NSF) Awards 2229876, 2320903, 2317184, and funds provided by the Department of Homeland Security, and by IBM.
This work is also supported in part by gifts from Intel and Activision.
Any opinions, findings, and conclusions or recommendations expressed in this publication are those of the authors and do not necessarily reflect the views of sponsors.

\bibliographystyle{plain}
\bibliography{ref}

\appendix
\section{Example of a Packet Parsed with Generic Photon Protocol Parser}
\label{appendix:example_photon_packet}
\begin{figure}[h]
\centering
\includegraphics[width=0.47\textwidth]{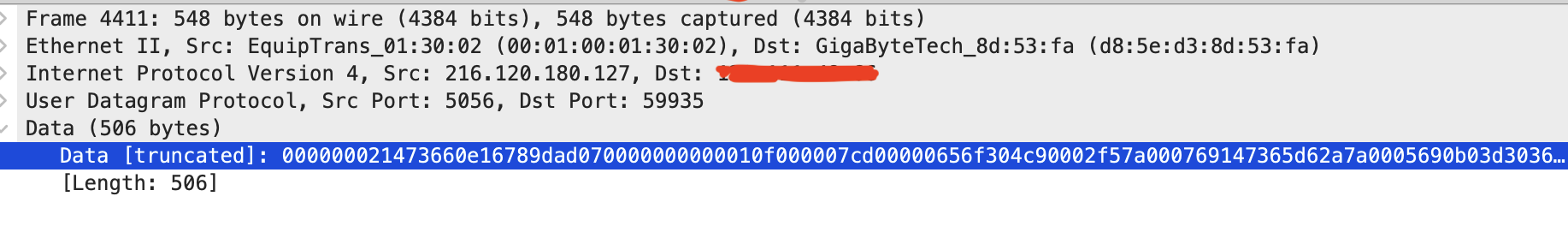}
\caption{Example of an unparsed motion data packet.}
\label{fig:unparsed_packet}
\end{figure}

In Figure~\ref{fig:unparsed_packet}, we show a Wireshark capture of a packet containing motion data sent by the Rec Room server (here, the IP address of the Rec Room server that sends motion data updates is 216.120.180.127). This motion data encodes an update of the movement of the avatar that is controlled by the victim user (and who is in the same virtual room as we are – the attacker). This motion update is used by our client to render the movement of the victim’s avatar. In our attack, we use this data to infer keystrokes. 

\begin{figure}[h]
\centering
\includegraphics[width=0.47\textwidth]{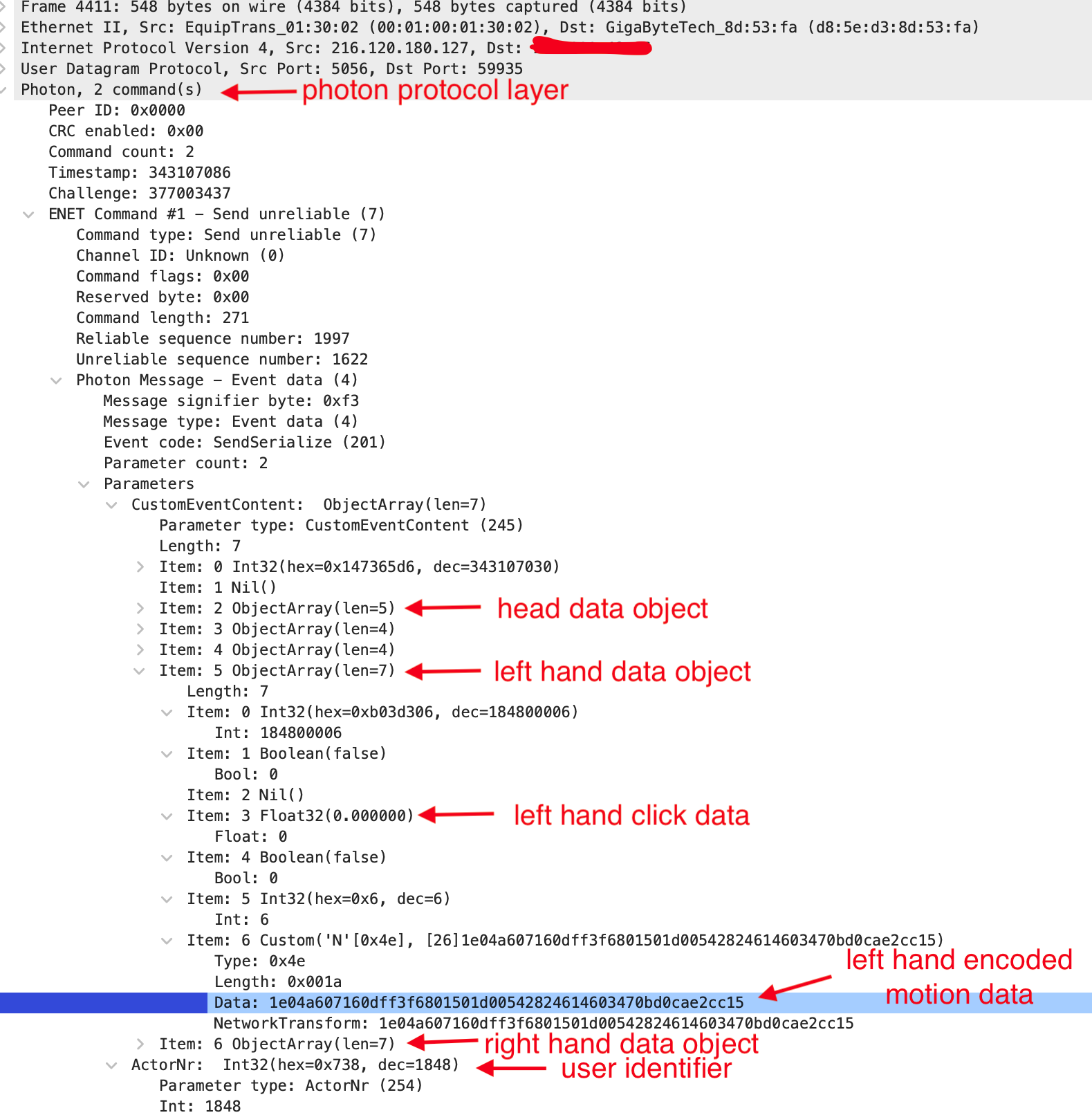}
\caption{Example of a parsed motion data packet from Rec Room.}
\label{fig:parsed_photon_packet_rec_room}
\end{figure}

\begin{figure}[h]
\centering
\includegraphics[width=0.47\textwidth]{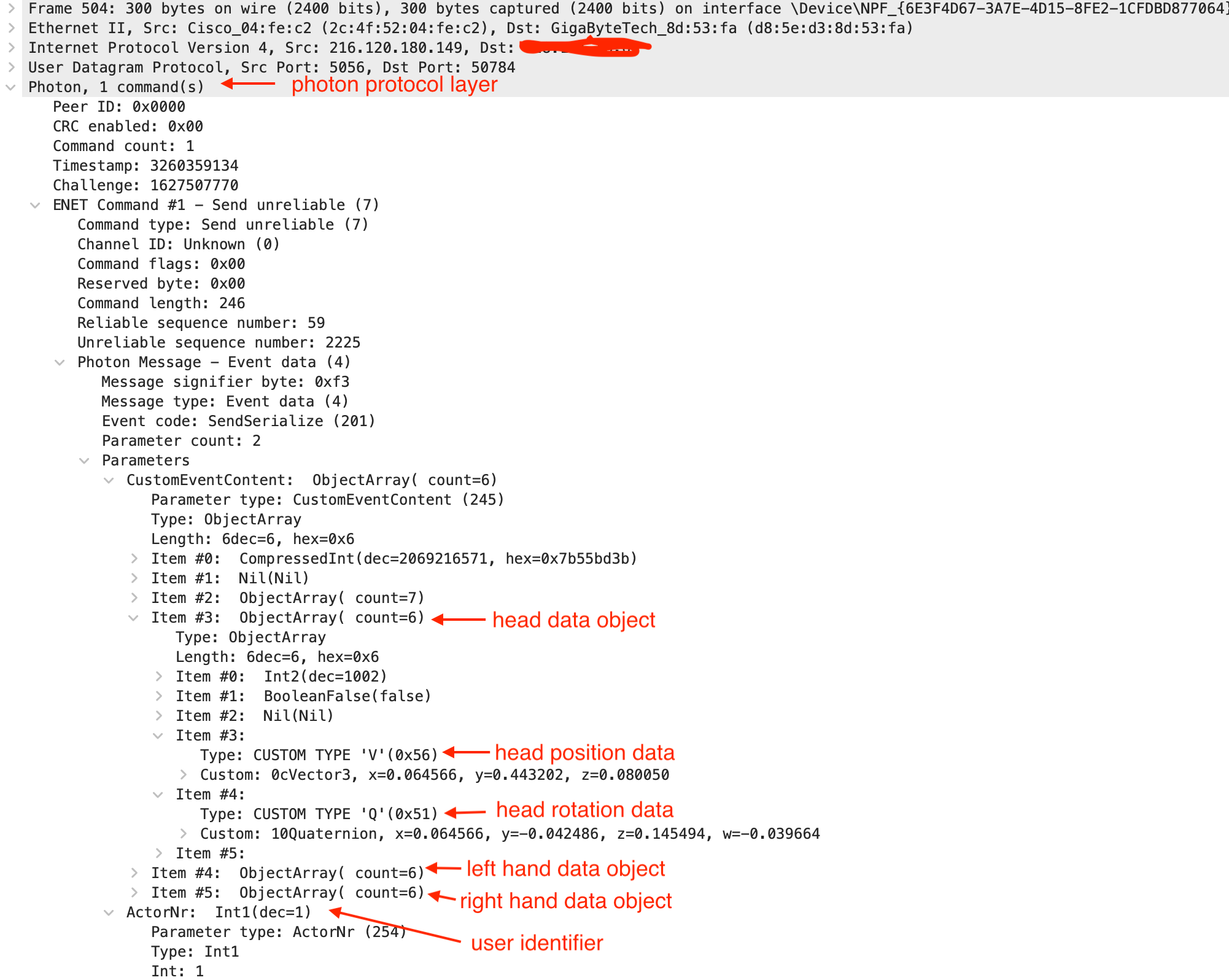}
\caption{Example of a parsed motion data packet from Galaxity.}
\label{fig:parsed_photon_packet_galaxity}
\end{figure}

\begin{figure}[h]
\centering
\includegraphics[width=0.47\textwidth]{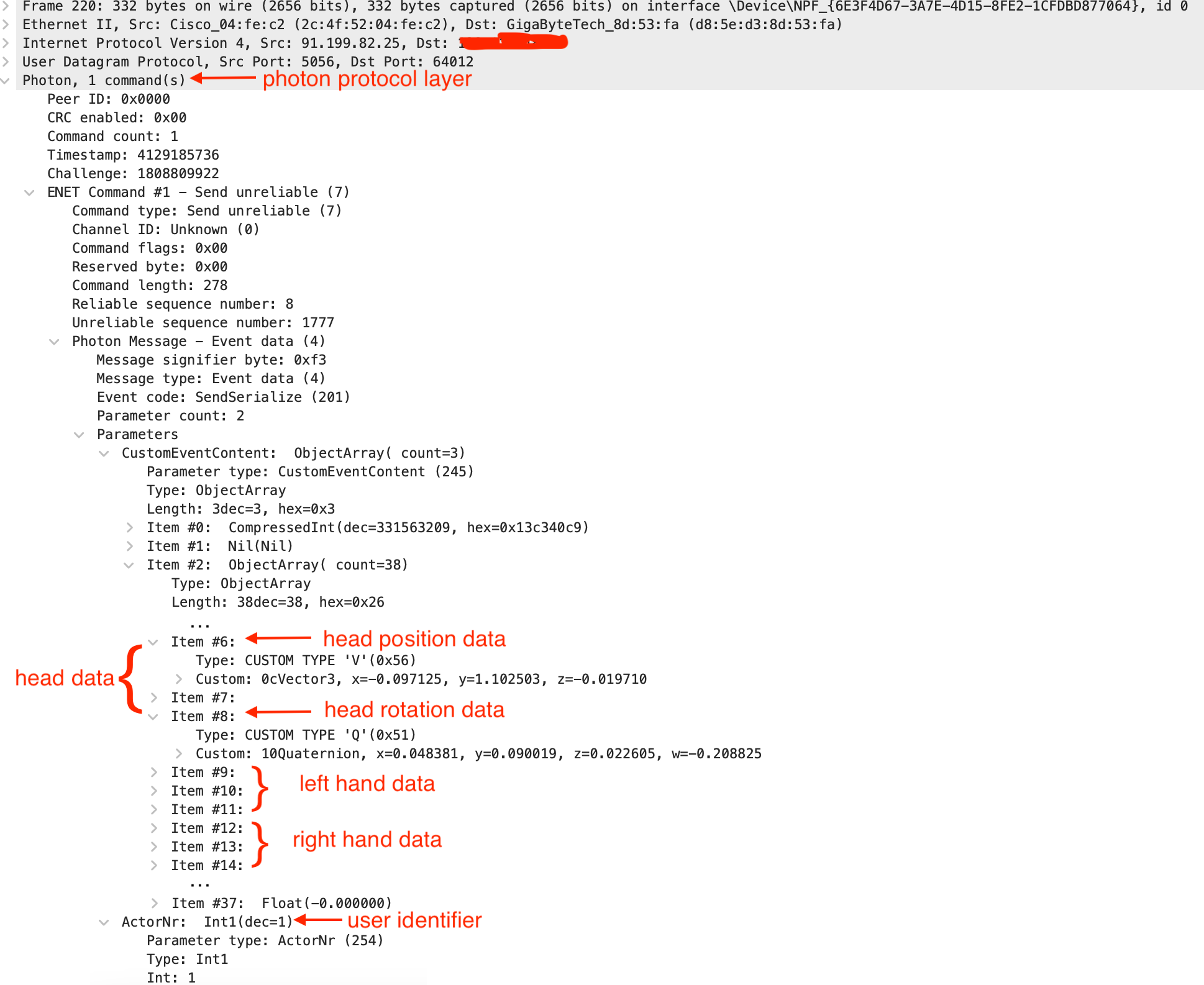}
\caption{Example of a parsed motion data packet from Sing Together: VR Karaoke.}
\label{fig:parsed_photon_packet_sing_together}
\end{figure}

\begin{figure}[h]
\centering
\includegraphics[width=0.47\textwidth]{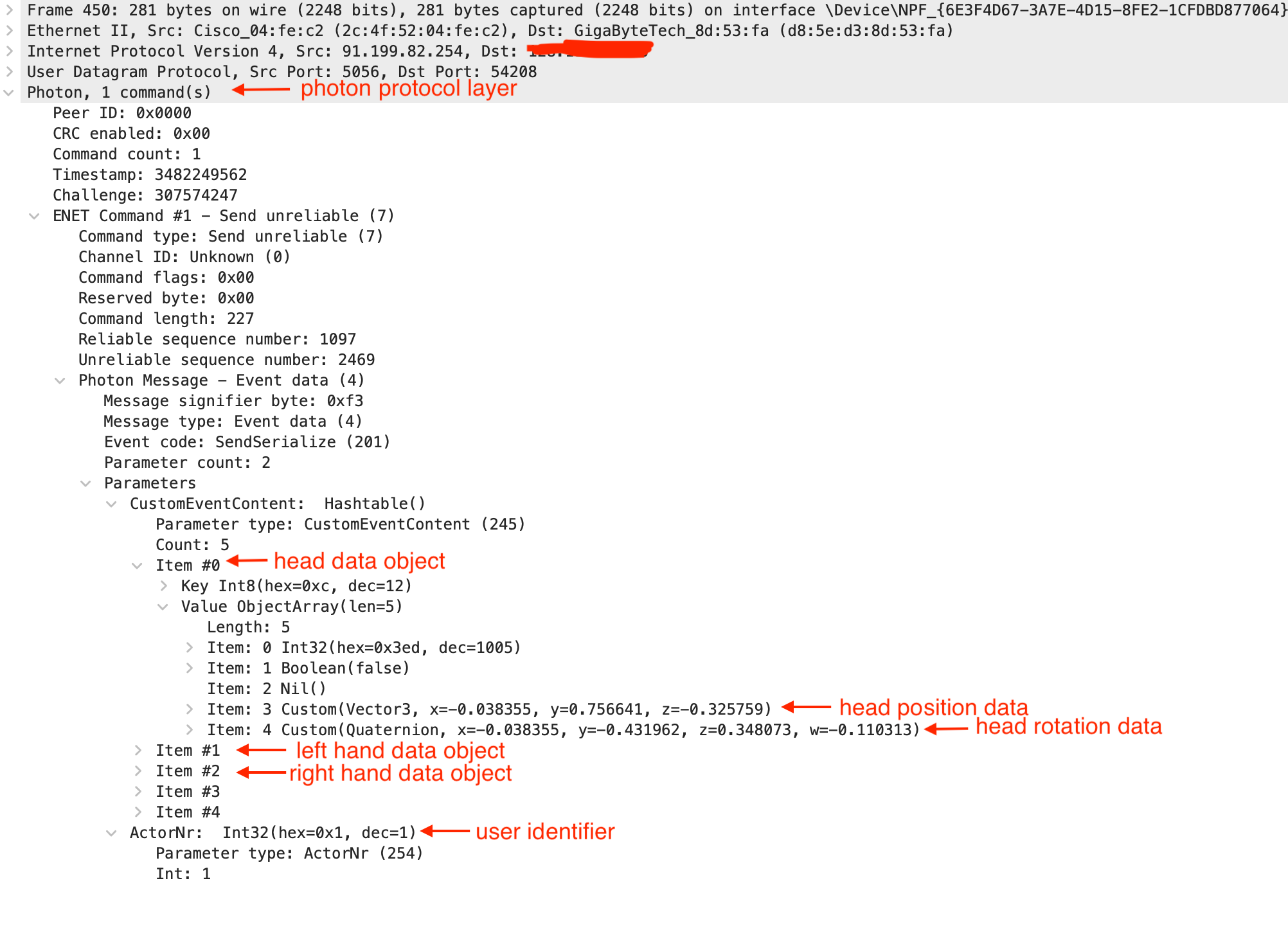}
\caption{Example of a parsed motion data packet from oVRshot.}
\label{fig:parsed_photon_packet_ovrshot}
\end{figure}

In Figure~\ref{fig:parsed_photon_packet_rec_room}, Figure~\ref{fig:parsed_photon_packet_galaxity}, Figure~\ref{fig:parsed_photon_packet_sing_together}, and Figure~\ref{fig:parsed_photon_packet_ovrshot}, we show how the packet payload is parsed by a generic parser for the Photon protocol (Section~\ref{subsec:field_extraction}). Photon is used to exchange objects between game clients and the Rec Room server. Note that, at this stage of parsing, we (the attacker) do not (yet) know the meaning (semantics) of specific fields of these objects, which are structured in different ways for different applications. However, the Photon parser can decode the raw packet payload into objects.

\section{Reverse-Engineering Custom Data Fields}
\label{reverse_engineer_appendix}

Rec Room is developed using the Unity game development engine.
Like all games developed using Unity, it is implemented mostly in C\#, however, the final game is usually emitted as a native code library using Unity's \textit{il2cpp} utility.
\textit{il2cpp} transpiles a C\# Unity project into an equivalent C++ project with a runtime providing most of the C\# standard library functionality.
For reverse-engineering purposes, this transpilation presents a major hurdle, as C\# applications generally include a plethora of high-level metadata of the target application, including function names, types, method signatures, classes, and many more.
The C++ binaries produced by \textit{il2cpp}, in contrast, do not include any such type information, and any included information can easily be removed (e.g. debug symbols) by application developers worried about reverse engineering.
However, since various C\# functionality (e.g., the ``Reflection'' APIs) requires fine-grained type information at runtime, \textit{il2cpp} produces a \textit{global-metadata.dat} file which contains the necessary type information omitted by the C++ compilation instead.
During gameplay, the \textit{il2cpp} runtime provides this type-information on-demand as the application requires it by loading and parsing this metadata file.

Various obfuscation and anti-tampering schemes exist for Unity developers to protect their games from reverse-engineering efforts as performed in this paper.
Notably, Epic Games offers EAC (Easy Anti-Cheat) to developers in the Unity store.
EAC is used by well-known games such as Fortnite, Apex Legends, HALO, etc., to prevent tampering with the game data by a malicious user in order to prevent cheating in these games.

RecRoom relies on EAC to prevent modification and/or introspection of game data at runtime.
To extract the motion data required for this project via dynamic analysis of the target application would require bypassing EAC's anti-cheat protections, since the required memory introspection capabilities can be used to implement various cheats like wall-hacks.
Such bypasses, while feasible, are highly guarded secrets of commercial cheat developers because any methods made public are generally quickly patched and mitigated.

Instead we focus on a fully static reverse-engineering approach to recover the necessary custom data-structures used by RecRoom to transmit motion data, events, player information, etc.
Photon Unity Networking (PUN), the networking library used by Rec Room, provides a common communication and serialization mechanism for a variety of game-related information, such as events, positions, rotations, entities, etc.
It also provides a common extension point for developers to send up to 256 arbitrary custom data types.
RecRoom uses these custom data types to implement more efficient custom encodings for motion data (among other things), e.g. Quaternion Compression and Quantization.
We identified the corresponding functionality in the RecRoom application by searching for custom types provided natively provided by PUN (namely \textit{Vector2/3} and \textit{Quaternion}).
Once the \textit{RegisterCustomType} function was identified, we cross-checked other call sites and found a function registering all custom types used by Rec Room.
Lastly, we reverse-engineered each custom handler to discover the internal structure for each custom type.

The now syntactically decoded packets are then used for the semantics-recovery process.

\section{Geometric Tests for Cursor and Key Measurement}
\paragraph{Cursor Measurement.}
\label{appendix:measure_cursor}
If all the measurements of the cursor are correct, we can verify them with the following geometric tests:

1) If we point the cursor at a vertical screen and move the hand along the measured forward direction of the cursor (i.e., which direction the cursor is pointing at), the cursor's reticle (projection of the cursor on the screen) should not move. 
Passing this test shows that the movement axis aligns with the cursor's real forward direction, and we have measured the cursor's orientation correctly. 
This test is illustrated in Figure~\ref{fig:geometric_test1}.

\begin{figure}[h]
\centering
\includegraphics[width=0.47\textwidth]{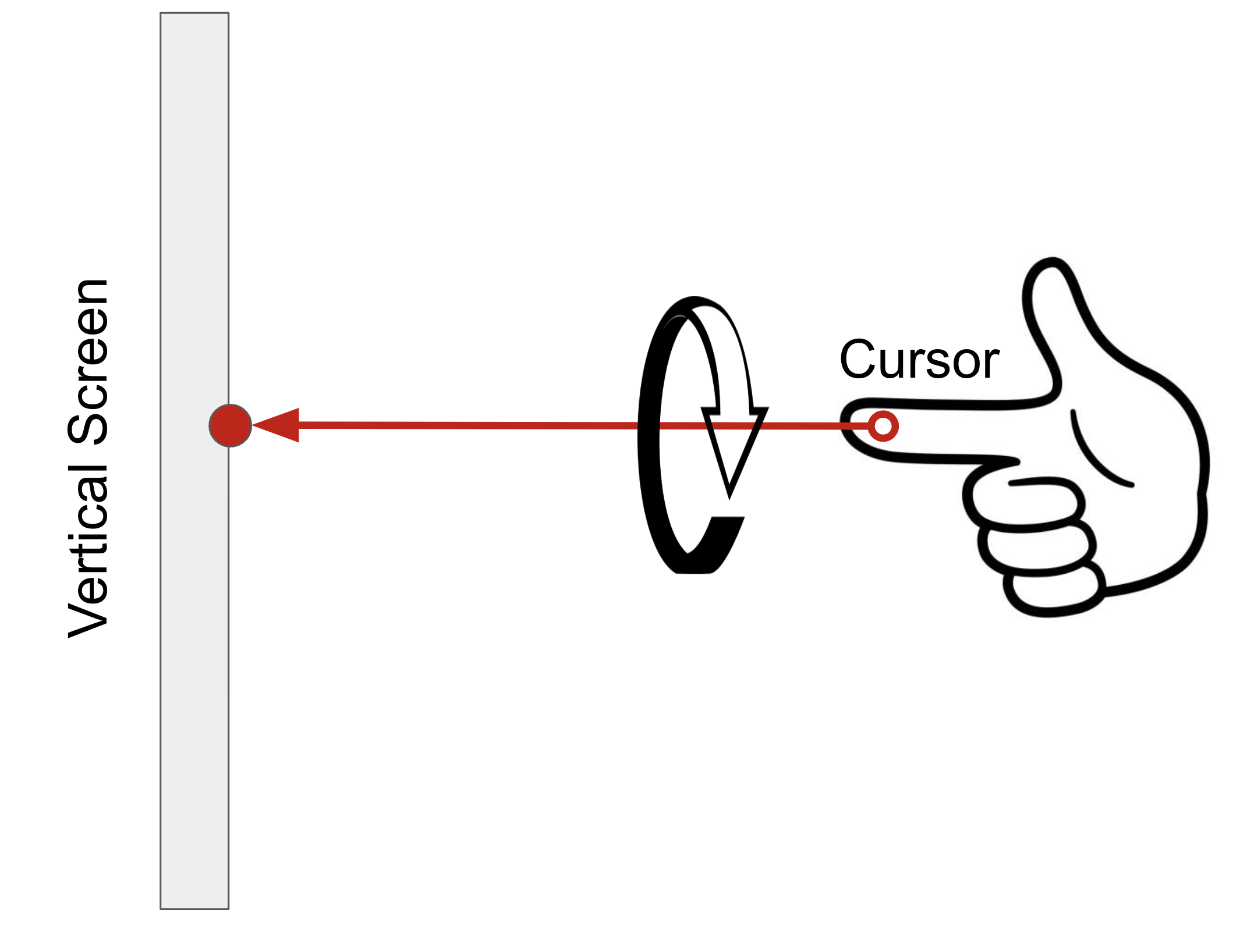}
\caption{Geometric test 2: test if the cursor position is correct by rotating the hand around the cursor's forward direction.}
\label{fig:geometric_test2}
\end{figure}

2) Once we measure the cursor's forward direction correctly, if we rotate the hand around the forward direction axis along the measured cursor's center point, the cursor's reticle should not move. 
Passing this test shows that the rotation axis crosses the measured cursor's center point and that we have measured the cursor's position correctly. 
This test is illustrated in Figure \ref{fig:geometric_test2}.

\begin{figure}[h]
\centering
\includegraphics[width=0.47\textwidth]{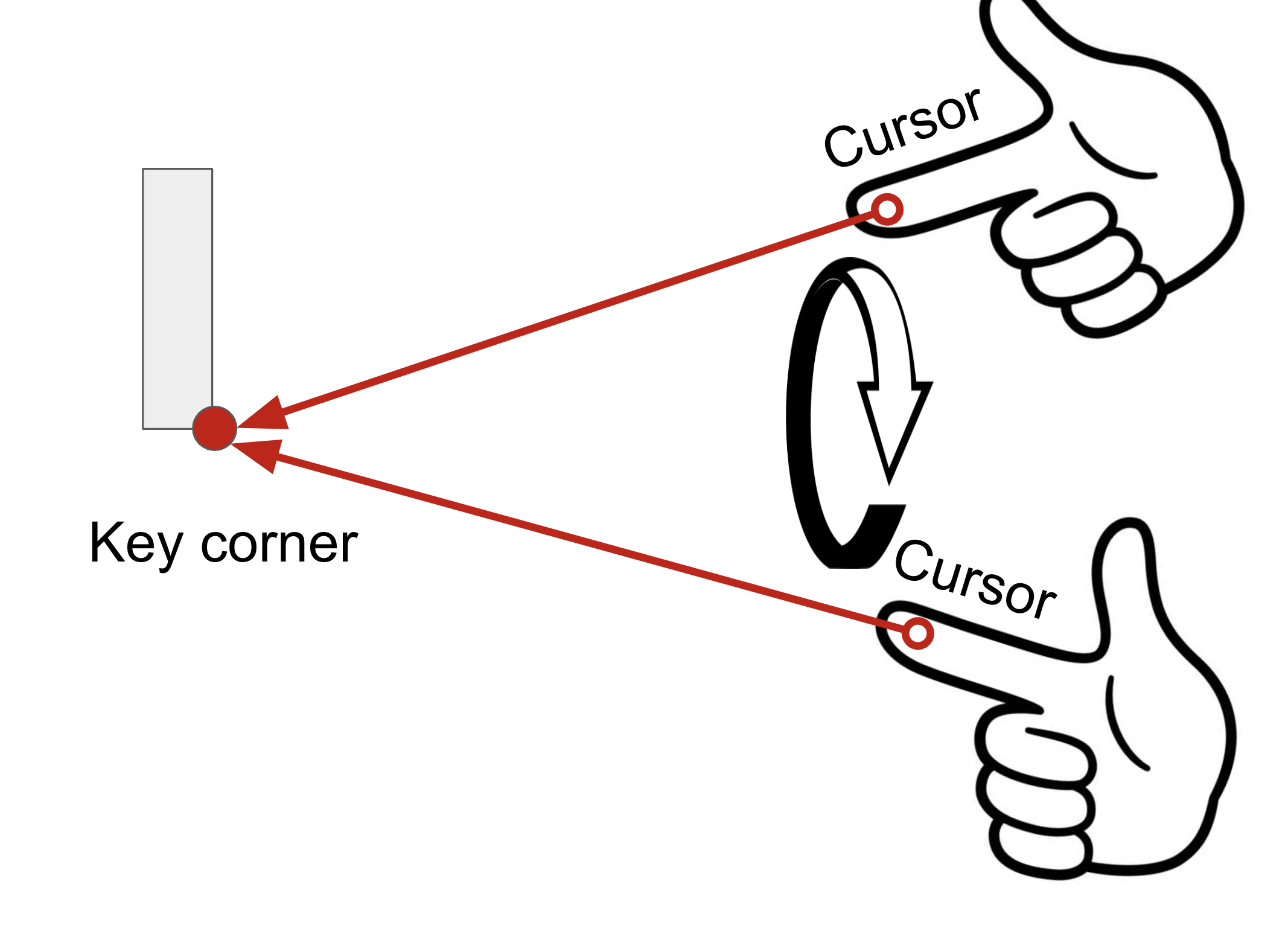}
\caption{Geometric test 3: test if the key corner measurement is correct by drawing a cone shape around the key corner with the cursor.}
\label{fig:geometric_test3}
\end{figure}

\paragraph{Key Measurement.}
\label{appendix:measure_keyboard}
If the measurement of a key's corner position is correct, we can verify it with the following geometric test:
1) Once we measure the cursor's \transform, if we can find an axis that crosses a key's corner, then move the cursor around this axis while pointing at the key's corner, the reticle should not move. 
Passing this test shows that we are able to triangulate the key's corner, and that we have measured the position of the key's corner correctly. 
This test is illustrated in Figure \ref{fig:geometric_test3}.

\section{Other Results for the Keylogging Attack}
\begin{figure}[h]
\centering
\includegraphics[width=0.5\textwidth]{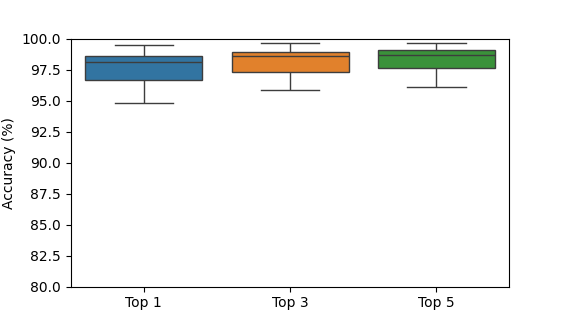}
\caption{The accuracy of the attack varies slightly among different participants due to their varying typing habits. }
\label{fig:acc_vs_participant}
\end{figure}

\paragraph{Attack Accuracy Across Participants.}
In Figure~\ref{fig:acc_vs_participant}, we can see the individual differences in the attack's accuracy. 
While the accuracy remains high across all participants (for top-$1$ accuracy, the minimum is 94.8\%), there are individual differences across participants (for top-$1$ accuracy, the median is 98.13\% and the maximum is 99.47\%). 
There may be many factors that contribute to this difference. 
For example, participants had different typing habits and positioning, as we did not want to put restrictions on the participants' typing process.
During the study, we observed that some participants leaned back and typed characters from a very far distance (some people even typed with awkward poses, positioning their hands above their shoulders), whereas others typed characters right in front of the keyboard. 
Therefore, it may be harder to predict keystrokes from those who were far from the keyboard similar to how farther keys were harder to predict as discussed in Section~\ref{sec:keyboard}). 
However, as the accuracy is still high across all participants, our attack is robust against individual differences.

\begin{figure}[h]
\centering
\includegraphics[width=0.5\textwidth]{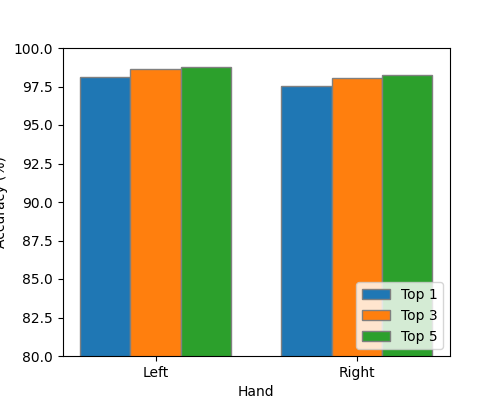}
\caption{The attack accuracy is not affected by the hand the victim uses to type. }
\label{fig:acc_vs_hand}
\end{figure}

\paragraph{Difference Across Hands.}
In Figure~\ref{fig:acc_vs_hand}, we show the attack's accuracy for both hands, which do not have noticeable difference (e.g., top-$1$ accuracy for the left hand is 98.1\%, and top-$1$ accuracy for the right hand is 97.51\%). 
This result is expected, as our attack uses hand motions to precisely calculate the keystroke, and the motions likely do not have a fundamental difference between the left hand and the right hand (e.g., they are likely positioned roughly the same distance from the keyboard). 

\section{Additional Applications}
\label{appendix:additional_app}
We recovered motion data from the network packets collected in the applications listed in Table~\ref{additional_app}.
\begin{table}[ht]
\begin{center}
\begin{tabular}{ll}
\toprule
          App Name   & Genre     \\
\midrule
Galaxity &social game  \\
Sing Together: VR Karaoke & music game  \\
oVRshot & shooter game \\

\bottomrule
\end{tabular}
\caption{List of additional applications on which we reproduced the steps to extract motion data.}
\label{additional_app}
\end{center}
\end{table}

\section{More Analysis for Attack on Partially Reverse Engineered Packets}
\label{mlother}

\paragraph{More Training Data Allows Higher Attack Accuracy.}

From the results demonstrated in Table~\ref{ml-result_data}, we found that when we train the machine learning classifier with more data, it can predict the keystrokes with higher accuracy. 
This highlights that if an adversary is able to collect more labeled typing motion from the victim, they can further improve the attack results.
\begin{table}[ht]
\begin{center}
\begin{tabular}{llll}
\toprule
Train Data & Top 1   & Top 3   & Top 5   \\
\midrule
20 Percent   & 30.16\% & 50.47\% & 63.01\% \\
40 Percent & 45.30\% & 67.81\% & 76.38\% \\
60 Percent & 57.14\% & 79.29\% & 86.54\% \\
80 Percent & 68.07\% & 85.96\% & 90.28\%\\
\bottomrule
\end{tabular}
\caption{More training data enables the model to better learn about decoding the data.}
\label{ml-result_data}
\end{center}
\end{table}

\paragraph{Secrets Typed with the Right Hand are More Vulnerable.}

From the result demonstrated in Table~\ref{ml-result_left_right}, we found that the data type with the right hand is slightly more vulnerable to our attack. 
We theorize that this is because most users predominantly use their right hand for typing, as evidenced by the data showing more than 70 percent of the clicks originate from the right hand. 
As a result, the model is trained on a larger volume of right-hand data, enhancing its generalizability when encountering new data for the testing samples when the victim types with the right hand.

\begin{table}[ht]
\begin{center}
\begin{tabular}{llll}
\toprule
          & Top 1   & Top 3   & Top 5   \\
\midrule
Left(Total)   & 66.73\% & 83.34\% & 88.91\% \\
Right(Total) & 72.31\% & 86.44\% & 90.70\% \\
\bottomrule
\end{tabular}
\caption{Secrets typed with the right hand are slightly more vulnerable to our machine learning models.}
\label{ml-result_left_right}
\end{center}
\end{table}

\paragraph{Performance of the Attack on Different Tasks.}

From the results demonstrated in Table~\ref{ml_result_task}, we found that the user's typing with just numbers can be inferred with a top-1 accuracy of 83.63\%. 
Typings with sentences can be inferred with an accuracy of 70.59\%, but password typing has a lower attack accuracy of 58.67\%.

We theorize that numbers are more vulnerable because they belong to fewer classes (only 10), causing each number to appear more frequently in the training dataset compared to the characters.
Sentences are also slightly more vulnerable compared to the main result.
This is because the distribution of characters within a sentence can be slightly imbalanced. 
Some characters might be seen more frequently than others, making them more vulnerable to our attack. 
However, since password typing can include any keys on the keyboard, some keys are rarely seen in the training data. These rarely-seen keys make passwords harder to attack.
\begin{table}[ht]
\begin{center}
\begin{tabular}{llll}
\toprule
          & Top 1   & Top 3   & Top 5   \\
\midrule
Numbers   & 83.63\% & 97.91\% & 100\% \\
Passwords & 58.67\% & 77.23\% & 86.61\% \\
Sentences & 70.59\% & 84.03\% & 92.44\% \\
\bottomrule
\end{tabular}
\caption{Attack accuracy for different typing tasks with partially reconstructed data.}
\label{ml_result_task}
\end{center}
\end{table}

\section{Survey}
\paragraph{Debrief.}
\label{debrief}
After the data collection was complete, we debriefed the participant on the real purpose of our study using the following scripts:

\begin{displayquote}
Thank you for your participation in this experiment. 
The goal of this study was to determine the vulnerabilities within the current VR typing systems and understand whether a malicious actor in the same virtual room can recover your keystrokes in a social VR app, which might be exploited to steal your private chat, password, or payment information. 
The result would be very helpful to further improve the general security of all Virtual Reality systems in the market. 

In this experiment, you were taught that the study was a study for typing in Virtual Reality. 
The reason behind not fully disclosing the study purpose was that we wanted you to complete the tasks without excessive caution so that your typing activities resemble a real-world scenario. 

Your participation is not only greatly appreciated by the researchers involved, but the data collected could possibly improve the security of Virtual Reality.  

Finally, we urge you not to discuss this study with anyone else who is currently participating or might participate at a future point in time. 
As you can certainly understand, we will not be able to examine the effectiveness of keystroke recovery in participants who know about the true purpose of the project beforehand. 
Thank you! 
\end{displayquote}

\paragraph{Survey Questions on User's Opinion of Keylogging Attack.}
\label{survey}
\begin{enumerate}
    \item What input method do you usually use to type in VR? 
    \begin{enumerate}
        \item Virtual Keyboard
        \item Voice inputs
        \item Traditional keyboard
        \item Hand gesture
        \item Eye-tracking
        \item Other:
    \end{enumerate}
    \item In which activities within VR games have you used typing? (select any that apply)
    \begin{enumerate}
        \item Private chat
        \item Email writing
        \item Browser search
        \item Password entry
        \item Payment information
        \item Other:
    \end{enumerate}
    \item Prior to this study, were you aware that any user in the same virtual room could potentially infer your keystrokes?
    \begin{enumerate}
        \item Yes
        \item No
    \end{enumerate}
    \item How concerned are you about the possibility of a malicious user inferring your keystrokes while in the same virtual room with you
    \begin{enumerate}
        \item Score 1: Not concerned at all
        \item Score 2: Slightly inclined to be concerned
        \item Score 3: Moderately disinclined to be concerned
        \item Score 4: Uncertain
        \item Score 5: Moderately inclined to be concerned
        \item Score 6: Strongly inclined to be concerned
        \item Score 7: Extremely concerned
    \end{enumerate}
\end{enumerate}

\end{document}